\documentstyle[amssymb,aps,eqsecnum,12pt,prc]{revtex}
\tightenlines

\begin{document}
\draft
\title{Radiative Corrections to Electron-Proton Scattering}
\author{L.C. Maximon}
\address{Physics Department, The George Washington University,\\
Washington, DC 20052, USA}
\author{J.A. Tjon}
\address{Institute for Theoretical
Physics, University of Utrecht, 3584 CC Utrecht, The Netherlands \\and\\
KVI, University of Groningen, 9747 AA Groningen, The Netherlands}
\date{4 February 2000}
\maketitle

\begin{abstract}
The radiative corrections to elastic electron-proton scattering are analyzed
in a hadronic model including the finite size of the nucleon. For initial
electron energies above 8 GeV and large scattering angles, the proton vertex
correction in this model increases by at least two percent the overall
factor by which the one-photon exchange (Rosenbluth) cross section must be
multiplied. The contribution of soft photon emission is calculated exactly.
Comparison is made with the generally used expressions previously obtained
by Mo and Tsai. Results are presented for some kinematics at high momentum
transfer.
\end{abstract}


\section{INTRODUCTION}

Electron scattering at intermediate and high energies has been one of the
most useful means of investigating nuclear structure for over forty years.
With the advent of CW accelerators and high resolution detectors such as
MAMI and TJNAF it has become clear that one must have an accurate estimate
of the radiative corrections if meaningful cross sections are to be obtained
from the experimental measurements. Depending on the experimental conditions
-- initial beam energy, momentum transfer, and detector resolution or
missing mass for the observed particles -- the radiative corrections can be
as large as $30\%$ of the uncorrected cross section. To obtain cross
sections which are accurate to 1\%, one must then know the radiative
correction to 3\%.

The theoretical expression for the radiative correction which has been used
in the analysis of almost all single arm elastic electron scattering
experiments with beam energies below approximately $25$ GeV (for which $W$
and $Z$ exchange are in general not significant) is that given originally by
Tsai \cite{tsai}, \cite{mo} in connection with experiments at Stanford, SLAC
and CEA. That work involved approximations that were both purely
mathematical (made in performing the integrations needed to evaluate the
inelastic cross section) and approximations denoted here as ``soft-photon
approximations'' that are directly related to the physics in that the effect
of proton structure was neglected; in considering the proton legs, only the
soft virtual (infrared) photon contribution is calculated exactly -
approximations are made in the hard virtual photon (non-infrared)
contribution. In particular, the proton structure is neglected by setting
the photon momentum square $k^{2}=0$ in the proton form factor, $F(k^{2})$,
thus simplifying the calculation considerably.

The purpose of the present paper is to study the radiative correction to
elastic electron-proton scattering including the internal structure of the
nucleon. For this we have considered a simple model in which the proton
current is taken to have the usual on-shell form. The model dependence of
the radiative correction is clearly an important question for the analysis
of electron scattering experiments at the 1\% level. This work is an initial
study to examine the size of internal structure effects.

The present calculation differs from that of Tsai \cite{tsai}, \cite{mo} in
three substantive aspects. First, we evaluate the inelastic cross section
(emission of soft real photons) without any approximation; the relevant
integrals have been given in closed form by t'Hooft and Veltman \cite{thooft}%
. In fact, these exact expressions are simpler in form than the approximate
ones given in \cite{tsai} and \cite{mo}. We note in particular that in the
limit of the target mass $M\rightarrow \infty ,$ corresponding to a static
Coulomb potential, we obtain exactly the result first given by Schwinger 
\cite{schwinger}. Second, in the evaluation of the contribution of the box
and crossed box diagrams to the elastic cross section we make a less drastic
approximation than that made in \cite{tsai}. Specifically, in the integrands
corresponding to the relevant matrix elements, $M_{2}$ and $M_{3}$ (Eqs. (%
\ref{box}) and (\ref{cbx})), we make a soft photon approximation (setting $%
k=0$ or $k=q$) in the {\it numerator} (as in \cite{tsai}), but not in the
denominators. Again, the required integrals (scalar four-point functions)
have been given in \cite{thooft}; the resulting expressions are again
considerably simpler than those obtained in \cite{tsai}, where the
soft-photon approximation is also made in the denominators of $M_{2}$ and $%
M_{3}$. Finally, in evaluating the proton vertex correction, we have made no
soft photon approximation for the virtual photon (as was done in \cite{tsai}%
) and have included form factors for the proton, taking the proton current
to be that indicated below in (\ref{cur}).

The organization of the paper is as follows: In Sec. II we discuss questions
concerning the electromagnetic nuclear current operator used in this
analysis. In Sec. III we give details of the calculation of the matrix
elements and cross section for elastic scattering, retaining terms of order $%
\alpha $ relative to the Rosenbluth (one photon exchange) cross section for
elastic scattering. Integrals needed for the evaluation of the various
matrix elements are written explicitly and expressed in closed form in terms
of Spence functions (dilogarithms). Details are given in the Appendices.

In Sec. IV we consider the inelastic cross section in detail; as with the
elastic cross section given in Sec. III, the result is expressed in closed
form in terms of Spence functions. In Sec. V we add the elastic and
inelastic cross sections, giving both an analytic expression and a numerical
evaluation of the radiative correction for various values of the pertinent
parameters, (initial beam energy, final electron detector resolution, and
target nucleus). We compare the values of the radiative correction
calculated here with those given in \cite{tsai} and \cite{mo}.

\section{Electromagnetic nucleon current operator}

We essentially follow in this paper the convention of Bj\"{o}rken and Drell 
\cite{bj}. The metric used is defined by

\begin{equation}
p_{i}\cdot p_{j}=\epsilon _{i}\epsilon _{j}-{\bf p}_{i}\cdot {\bf p}_{j}
\label{A1}
\end{equation}

Further, $\alpha =e^{2}/4\pi =1/137.036$; $m$ is the electron rest mass; $M$
is the target nucleus rest mass; $Z$ the charge of the target nucleus; $%
\kappa $ the anomalous magnetic moment of the proton; $p_{1}$ and $p_{3}$
the initial and final electron four-momenta, respectively; $p_{2}$ and $%
p_{4} $ the initial and final target nucleus four-momenta, respectively; $%
q=p_{1}-p_{3}=p_{4}-p_{2}$ is the four-momentum transfer to the target
nucleus for elastic scattering. It will prove useful to define, in addition, 
\begin{equation}
\rho =p_{4}+p_{2},\,\,\,\rho _{m}=p_{1}+p_{3},
\end{equation}
from which $\rho ^{2}=-q^{2}+4M^{2}$ and $\rho _{m}^{2}=-q^{2}+4m^{2}$.
Further, we define 
\begin{eqnarray}
x &=&(\rho +\rho _{1})\diagup (\rho -\rho _{1})=(\rho +\rho _{1})^{2}\diagup
4M^{2}  \nonumber \\
x_{m} &=&(\rho _{m}+\rho _{1})\diagup (\rho _{m}-\rho _{1})=(\rho _{m}+\rho
_{1})^{2}\diagup 4m^{2}
\end{eqnarray}
with $\rho _{1}^{2}=-q^{2}$. In the lab system we have: $p_{1}=(\epsilon
_{1},{\bf p}_{1});\,\,\,p_{3}=(\epsilon _{3},{\bf p}_{3});\,\,\,p_{2}=(M,0);%
\,\,\,p_{4}=(M+\omega ,{\bf q);\,\,\,}\omega =-q^{2}/2M$.

With the aim of presenting expressions which correspond to the experimental
conditions of high energy electron scattering, we neglect, in the final
expressions given in this paper, terms of relative orders $%
m^{2}/\epsilon^{2},$ $m^{2}/(-q^{2})$, and $m^{2}/M^{2}$. Neglect of these
terms defines our high energy approximation. No assumption is made, however,
with regard to the magnitudes of $M/\epsilon _{1},$ $M/\epsilon _{3},$ or $%
M^{2}/(-q^{2}).$

At low momentum transfer the internal structure of the nucleon can safely be
neglected in the determination of the radiative corrections in
electron-nucleus scattering. However, with increasing energies and momenta
this is in general no longer the case. One of the objectives of this paper
is to investigate this in a model for the e.m. interaction of a
non-pointlike nucleon. The most general e.m. off-shell nucleon vertex can be
characterized by 6 invariant functions \cite{NK,TT}. As the most simple
model we may consider a vector dominance-like model for the nucleon current,
characterized by only two form factors which depend only on the
four-momentum square of the photon. It is given by 
\begin{equation}
\Gamma _{\mu }=F_{1}(q^{2})\gamma _{\mu }+\kappa F_{2}(q^{2})\frac{i\sigma
_{\mu \nu }q^{\nu }}{2M},  \label{cur}
\end{equation}
where the form factors $F_{1}(q^{2})$ and $F_{2}(q^{2})$ are taken to have a
monopole or dipole form: 
\begin{equation}
F_{1}(q^{2})=F_{2}(q^{2})=\left( \frac{-\Lambda ^{2}}{q^{2}-\Lambda ^{2}}%
\right) ^{n},\,\,\,\,\,\,\,\,\,\,\,\,\,\,n=1\,\text{\thinspace or\thinspace
2 }  \label{ff}
\end{equation}
with $\Lambda $ being a constant of the order of $1\,\,$GeV/$c.$
Furthermore, $q=p^{\prime }-p$, $p$ and $p^{\prime }$ being the momentum of
the initial and final nucleon. Although the quantitative predictions of the
radiative corrections are expected in general to be dependent on the details
of the nucleon model assumed, one should already be able to see most of the
salient features in the present model study. In particular, identifying
regions in phase space where the finite size of the nucleon may play an
important role in the size of radiative corrections can be important. In
this way one may hope to get some feeling on the reliability of neglecting
the internal structure of the nucleon as is usually done. The present study
is intended as a first exploration of the sensitivity on the non-pointlike
nature of the e.m. hadronic current. As in \cite{tsai}, although we are
primarily interested in electron-proton scattering, the radiative
corrections studied here can also be applied to electron-nucleus scattering,
with appropriate changes in $F_{1},$ $F_{2},\,\,\kappa ,$ and $M$. However,
even in the case of electron-proton scattering, the factor $Z$ is convenient
for identifying the contributions from the various diagrams.

It should be noted that the dressed vertex function, $\widetilde{\Lambda }%
_{\mu },$ with Eq. (\ref{cur}) as e.m. current operator containing the form
factors $F_{n}$, satisfies a Ward-Takahashi identity 
\begin{equation}
q^{\mu }\widetilde{\Lambda }_{\mu }=F_{1}(q^{2})\left[ S^{-1}(p^{\prime
})-S^{-1}(p)\right]  \label{wt}
\end{equation}
where $S$ is the dressed nucleon propagator. As a direct consequence of (\ref
{wt}), one gets for on-mass-shell nucleons, the current conservation 
\begin{equation}
q^{\mu }<p^{\prime }|\widetilde{\Lambda }_{\mu }|p>=0.
\end{equation}
Obviously, the radiative corrections will in general be sensitive to the
choice of the e.m. current. Although interesting in its own right we will
not address in this paper the issue of the dependence of the predictions on
this ambiguity.

In the study of radiative corrections we may distinguish between the elastic
and inelastic contributions, the latter being the real soft photon emission
processes from both the electron and hadron. The elastic electron cross
section can be determined immediately from the total scattering amplitude $%
{\cal M}$ through the well-known expression

\begin{eqnarray}
d\sigma &=&\frac{mM}{\sqrt{(p_{1}\cdot p_{2})^{2}-m^{2}M^{2}}}%
\sum_{spins}\int \left| {\cal M}\right| ^{2}(2\pi )^{4}\delta
^{4}(p_{4}+p_{3}-p_{2}-p_{1})  \nonumber \\
&&\,\,\,\,\,\,\,\,\,\,\,\,\,\,\,\,\,\,\,\,\,\,\,\,\,\,\,\,\,\,\,\,\,\,\,\,\,%
\,\,\,\,\,\,\,\,\,\,\,\,\,\,\,\,\,\,\,\,\times \,\frac{m\,d^{3}p_{3}}{(2\pi
)^{3}\epsilon _{3}}\,\,\,\frac{M\,d^{3}p_{4}}{(2\pi )^{3}\epsilon _{4}},
\label{xs}
\end{eqnarray}
For single-arm experiments with unpolarized electrons in which the final
proton is not observed, $d\sigma $ must be averaged over initial spins,
summed over final spins, and integrated over the final proton four-momentum.
Up to order $\alpha ^{2}$ we have for the total scattering amplitude 
\begin{equation}
{\cal M}=\sum_{n=1}^{6}M_{n},  \label{total}
\end{equation}
where the various terms correspond to the Feynman graph contributions shown
in Fig.\ \ref{fig1}. $M_{1}$ is the matrix element for the one-photon
exchange diagram
\begin{equation}
M_{1}=Ze^{2}\overline{u}(p_{3})\gamma _{\mu }u(p_{1})\frac{(-i)}{%
q^{2}+i\epsilon }\overline{u}(p_{4})\Gamma ^{\mu }(q^{2})u(p_{2}).
\label{ope}
\end{equation}
Its square gives, for high energy electrons, the Rosenbluth cross section: 
\begin{equation}
\frac{d\sigma _{0}}{d\Omega }=\frac{\alpha ^{2}\cos ^{2}\frac{\theta }{2}%
\left[ \left( F_{1}^{2}-\frac{\kappa ^{2}q^{2}}{4M^{2}}F_{2}^{2}\right) -%
\frac{q^{2}}{2M^{2}}\left( F_{1}+\kappa F_{2}\right) ^{2}\tan ^{2}\frac{%
\theta }{2}\right] }{4\epsilon _{1}^{2}\eta \sin ^{4}\left( \theta /2\right) 
},  \label{rosen}
\end{equation}
where $\,\eta $ is the lab system recoil factor: For $\epsilon
_{1}>>m,\epsilon _{3}>>m,\,\,\,\,\,\,\,\eta \cong \epsilon _{1}/\epsilon
_{3}\cong 1+(\epsilon _{1}/M)(1-\cos \theta )$ with $\theta $ being the
electron scattering angle. We note, in particular, that $1\leq \eta \leq x$.
Furthermore, $M_{2}$ and $M_{3}$ are the matrix elements for the box and
crossed box (two-photon exchange) diagrams. $M_{4}$ is the vacuum
polarization diagram (only an electron-positron loop is indicated in the
figure, but the contribution from higher mass lepton loops can be included
without difficulty- see (\ref{vp2}) -(\ref{vp4})). $M_{5}$ is the electron
vertex correction, and $M_{6}$ is the proton vertex correction. For
completeness, we list in Appendix A the explicit expressions for the various
Feynman diagrams shown in Figs.\ \ref{fig1} and \ref{fig2}.

\section{Elastic Cross Section}

To evaluate the various one-loop corrections to Eq.~(\ref{total}) some
tedious algebra has to be carried out. We outline the procedure used to
evaluate the matrix elements needed for the radiative correction to the
elastic cross section, $M_{2}$ through $M_{6}$, (Eqs. (\ref{A3})-(\ref{vp3}%
)).

\subsection{Proton vertex correction}

We begin with the matrix element for the proton vertex correction, $M_{6}$ ,
given by Eqs. (\ref{A4}) and (\ref{A6}); the much simpler matrix element for
the electron vertex correction, $M_{5}$, (Eq. (\ref{A3})), can be deduced
quite easily from that. In (\ref{A6}), each of the three $\Gamma $'s, given
by Eq. (\ref{cur}), contains a term with $\gamma _{\mu }$ (which we denote
by $g$) and a term with $\sigma _{\mu \nu }$ (which we denote by $s$). The
proton vertex correction $\Lambda ^{\mu }(p_{4},p_{2})$ then consists of
eight terms, which we represent symbolically by $ggg,gsg,gss,$ etc. As may
be seen after rationalizing the propagators, the $k$ dependence of the
numerators for $ggg,gsg,...$ is such that there are at most four factors of
the form $k\!\!\!/ $. Moreover, the terms with three or four factors $%
k\!\!\!/$ may, with only a minimum of algebra, be written so that two of
these factors are adjacent, giving $k\!\!\!/k\!\!\!/=k^{2}$. Although the
calculation can equally well be carried out with $F_{1}$ and $F_{2}$
distinct functions, we assume $F_{1}=$ $F_{2}=F$, which simplifies the
algebra. The terms $ggg,gsg,...$ can then be expressed in terms of the
integrals

\begin{equation}
\{I_{0};I_{\mu };I_{\mu \nu };J_{0};J_{\mu };J_{\mu \nu };K_{0}\}=\int \frac{%
d^{4}k}{(2\pi )^{4}}\,F^{2}(k^{2})\{1;k_{\mu };k_{\mu }k_{\nu };k^{2};k_{\mu
}k^{2};k_{\mu }k_{\nu }k^{2};(k^{2})^{2}\}/D(\lambda ^{2})  \label{3.1}
\end{equation}
where 
\begin{equation}
D(\lambda ^{2})=(k^{2}-\lambda ^{2}+i\epsilon )(k^{2}-2k\cdot
p_{2}+i\epsilon )(k^{2}-2k\cdot p_{4}+i\epsilon )  \label{3.2}
\end{equation}
For form factors having the form given in (\ref{ff}), the integrals in (\ref
{3.1}) could all be evaluated as indicated for three-point functions in \cite
{thooft}, Sec.5, and \cite{passarino}, Appendix E. However, in the interest
of obtaining a relatively compact analytic expression in closed form, we
have used an alternative procedure. As given here in Appendix B, the
integrals may be expressed in terms of their moments, defined by Eqs. (\ref
{g1h1})-(\ref{g22h22}) and (\ref{B13}). After straightforward though
somewhat tedious algebra, the terms $ggg,gsg,...$ are then expressed in
terms of these moments. Next, for form factors of the form given in (\ref{ff}%
), we show that all of the moments may be expressed in terms of three
functions, $\phi _{k}$, which obey a three-term inhomogenous recursion, and
this is used for their evaluation. Finally, we note from (\ref{ggg2}) - (\ref
{Sm1}) that the terms $ggg,gsg,...$ may be usefully grouped by writing them
in the form 
\begin{equation}
(g+s)g(g+s)=F(q^{2})\left[ G_{1}(q^{2})\gamma _{\mu }+G_{2}(q^{2})\frac{%
i\sigma _{\mu \nu }q^{\nu }}{2M}\right]  \label{gpsggps}
\end{equation}
and 
\begin{equation}
(g+s)s(g+s)=\kappa F(q^{2})\left[ X_{1}(q^{2})\gamma _{\mu }+X_{2}(q^{2})%
\frac{i\sigma _{\mu \nu }q^{\nu }}{2M}\right]  \label{gpssgps}
\end{equation}

We note in the expressions for $ggg$, $gsg$,... in Appendix B that the
infrared divergent terms are all contained solely within $ggg$ and $gsg$.
These are the terms with a factor $\phi _{1}(\lambda ^{2})$ in (\ref{ggg2})
and (\ref{gsg2}). Since these are precisely the terms which are retained in
the proton vertex correction in \cite{tsai}, we separate them for the
purpose of comparison with that work, writing $M_{6}$ in the form\hspace{0in}
\begin{equation}
M_{6}=M_{6}^{(0)}+M_{6}^{(1)}  \label{M0M1}
\end{equation}
where 
\begin{equation}
M_{6}^{(0)}=-\frac{\alpha Z^{2}}{2\pi }(2M^{2}-q^{2})\phi _{1}(\lambda
^{2})M_{1}
\end{equation}
The function $\phi _{1}(\lambda ^{2})$, defined by (\ref{B20}), (\ref{phik})
and (\ref{C0l}) and evaluated in (\ref{phi1}) and (\ref{phi1lam}), is simply
related to the function $K(p_{2},p_{4})$ defined in \cite{tsai} by 
\[
K(p_{i},p_{j})=\frac{2p_{i}\cdot p_{j}}{-i\pi ^{2}}\int \frac{d^{4}k}{%
(k^{2}-\lambda ^{2}+i\epsilon )(k^{2}-2k\cdot p_{i}+i\epsilon
)(k^{2}-2k\cdot p_{j}+i\epsilon )} 
\]
{\it viz}., 
\begin{equation}
K(p_{2},p_{4})=2p_{2}\cdot p_{4}\phi _{1}(\lambda ^{2})  \label{Kp2p4}
\end{equation}

\subsection{Proton self energy correction}

We next consider the contribution of the proton self energy diagrams. It is
given by $\Sigma ^{^{\prime }}$ where 
\begin{equation}
\Sigma ^{^{\prime }}=\frac{1}{4} Tr \left[ \frac{\partial \Sigma }{\partial
p\!\!\!/} \right]  \label{sigprim}
\end{equation}
in which $\Sigma $ is the lowest order self-energy contribution 
\begin{equation}
\Sigma =-ie^{2}\int \frac{d^{4}k}{(2\pi )^{4}}\frac{1}{k^{2}-\lambda
^{2}+i\epsilon }\Gamma ^{\nu }(k)\frac{1}{(p\!\!\!/-k\!\!\!/-M+i\epsilon )}%
\Gamma _{\nu }(k)  \label{sig}
\end{equation}
Using the Ward-Takahashi identity Eq. (\ref{wt}) we find that 
\[
\Sigma ^{^{\prime }}=G_{1}(0) 
\]
where $G_{1}(q^{2})$ is the coefficient of $\gamma _{\mu }$ in $ggg$ (\ref
{ggg2}). Explicitly, $G_{1}(0)$ is given by Eqs.(\ref{B28p}) - (\ref
{phi10Lam}) and (\ref{phi10lam}). The addition of this contribution to the
lowest order vertex correction modifies the expressions for $(g+s)g(g+s)$
and $(g+s)s(g+s)$ given above in (\ref{gpsggps}) and (\ref{gpssgps}) so that
we now have 
\begin{equation}
\overline{(g+s)g(g+s)}=F(q^{2})\left[ \left( G_{1}(q^{2})-G_{1}(0)\right)
\gamma _{\mu }+G_{2}(q^{2})\frac{i\sigma _{\mu \nu }q^{\nu }}{2M}\right]
\end{equation}
and 
\begin{equation}
\overline{(g+s)s(g+s)}=\kappa F(q^{2})\left[ X_{1}(q^{2})\gamma _{\mu
}+\left( X_{2}(q^{2})-G_{1}(0)\right) \frac{i\sigma _{\mu \nu }q^{\nu }}{2M}%
\right]
\end{equation}
We then have, for the matrix element including self energy diagrams, 
\begin{equation}
\overline{M}_{6}=\overline{M}_{6}^{(0)}+\overline{M}_{6}^{(1)}  \label{M6bar}
\end{equation}
where 
\begin{equation}
\overline{M}_{6}^{(0)}=\frac{\alpha Z^{2}}{2\pi }\left[
-K(p_{2},p_{4})+K(p_{2},p_{2})\right] M_{1}  \label{M6bar0}
\end{equation}
which is the expression given in \cite{tsai}, eq. (II.12). The infrared
divergent part of these terms is cancelled exactly by the infrared divergent
terms in the inelastic cross section. The contribution of the matrix element 
$M_{6}^{(1)}$, which depends on the proton form factor, will be considered
after we write the electron vertex and box diagram corrections.

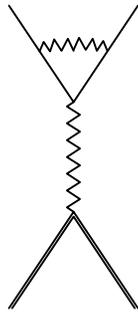
\subsection{Electron vertex correction}

The electron vertex correction, $M_{5}$, may be obtained directly from the
proton vertex correction, $M_{6}$. The expression $\Lambda _{\mu
}(p_{3},p_{1})$, Eq. (\ref{A5}), follows from $\Lambda ^{\mu }(p_{4},p_{2})$
if we retain only the term $ggg$, set $F=1$, replace $p_{2},$ $p_{4}$ and $M$
by $p_{1},$ $p_{3}$ and $m$, and, after performing the integrations, take
the limit $\Lambda \rightarrow \infty $ (note Eq.(\ref{ff})). Note that $%
\rho =p_{4}+p_{2}$ is then replaced by $\rho _{m}=p_{3}+p_{1}$, and $x,$
defined in section II, is replaced by $x_{m}$. Details are given in Appendix
C. We find

\begin{equation}
ggg=G_{1}^{(e)}(q^{2})\gamma _{\mu }+G_{2}^{(e)}(q^{2})\frac{i\sigma _{\mu
\nu }q^{\nu }}{2m}
\end{equation}
where 
\begin{equation}
G_{1}^{(e)}(q^{2})=\frac{\alpha }{2\pi }\left\{ -K(p_{1},p_{3})+\left( \frac{%
3\rho _{1}^{2}+8m^{2}}{2\rho _{m}\rho _{1}}\right) \ln x_{m}+\frac{1}{4}+%
\frac{1}{2}\ln \left( \frac{\Lambda ^{2}}{m^{2}}\right) \right\}
\end{equation}
\begin{equation}
G_{2}^{(e)}(q^{2})=\frac{\alpha }{2\pi }\left\{ \frac{2m^{2}}{\rho _{m}\rho
_{1}}\ln x_{m}\right\}
\end{equation}
Adding the contribution of the electron self energy diagrams gives 
\begin{equation}
\overline{ggg}=\left( G_{1}^{(e)}(q^{2})-G_{1}^{(e)}(0)\right) \gamma _{\mu
}+G_{2}^{(e)}(q^{2})\frac{i\sigma _{\mu \nu }q^{\nu }}{2m}
\end{equation}
from which 
\begin{equation}
\overline{ggg}=\frac{\alpha }{2\pi }\left\{ \left(
-K(p_{1},p_{3})+K(p_{1},p_{1})+\frac{3\rho _{1}^{2}+8m^{2}}{2\rho _{m}\rho
_{1}}\ln x_{m}-2\right) \gamma _{\mu }+\left( \frac{2m^{2}}{\rho _{m}\rho
_{1}}\ln x\right) \frac{i\sigma _{\mu \nu }q^{\nu }}{2m}\right\}
\label{elvtx}
\end{equation}

For large momentum transfers, $-q^{2}\gg m^{2}$, this reduces to 
\begin{equation}
\overline{ggg}=\frac{\alpha }{2\pi }\left\{ -K(p_{1},p_{3})+K(p_{1},p_{1})+%
\frac{3}{2}\ln \left( \frac{-q^{2}}{m^{2}}\right) -2\right\} \gamma _{\mu }
\end{equation}
Comparing (\ref{ope}) and (\ref{A3}), we then have, for $-q^{2}\gg m^{2}$,
\begin{equation}
\overline{M}_{5}=\frac{\alpha }{2\pi }\left\{ -K(p_{1},p_{3})+K(p_{1},p_{1})+%
\frac{3}{2}\ln \left( \frac{-q^{2}}{m^{2}}\right) -2\right\} M_{1}
\label{M5bar}
\end{equation}
which is the expression given in \cite{tsai}, eq.(II.5). We note that the
infrared divergence is contained entirely within the terms $%
-K(p_{1},p_{3})+K(p_{1},p_{1}).$ The infrared divergent part of these terms
is cancelled exactly by the infrared divergent terms in the inelastic cross
section.

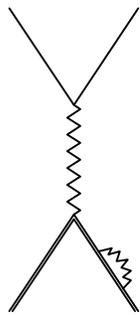
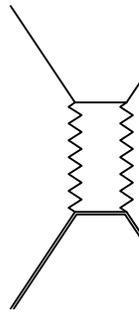
\subsection{Box and crossed-box diagrams}

The matrix elements for the box and crossed-box diagrams, $M_{2}$ and $M_{3}$%
, are given in (\ref{box}) and (\ref{cbx}). After rationalizing the
propagators, the required integrals can, for form factors of the form (\ref
{ff}), all be written in terms of four-point functions; in principle they
can be evaluated using \cite{thooft}, Sec. 6, and \cite{passarino}, Appendix
E. For the present work, however, we have chosen to evaluate these matrix
elements in an approximate manner, but one which is less drastic than that
employed in \cite{tsai}. We note first in $M_{2}$ and $M_{3}$ that the
integrands in $M_{2}$ and $M_{3}$ have two infrared divergent factors, $%
\,[(k^{2}-\lambda ^{2}+i\epsilon )((k-q)^{2}-\lambda ^{2}+i\epsilon )]^{-1}$%
. The integrands are thus peaked when either of the two exchanged photons is
soft, and become divergent when $k\rightarrow 0$ or when $k\rightarrow q$.
We therefore evaluate the numerators in $M_{2}$ and $M_{3}$ at these two
points but make no changes to the denominators. A simple calculation shows
that in fact that each of the numerators has the same value for $k=0$ as for 
$k=q$, {\it viz}., $4ip_{1}\cdot p_{2}\,q^{2}M_{1}$ in the case of $M_{2}$
and $4ip_{3}\cdot p_{2}\,q^{2}M_{1}$ in the case of $M_{3}$. We then take
this factor outside of the integral and are left with a scalar four-point
function to evaluate. The result has been given in \cite{thooft}, Sec. 6 and
Appendix E (b) and is expressed simply in terms of logarithms: 
\begin{equation}
M_{2}=-\frac{\alpha Z}{\pi }\frac{\epsilon _{1}}{|{\bf {\text{p}}_{1}|}}\ln
\left( \frac{\epsilon _{1}+|{\bf {\text{p}}_{1}|}}{m}\right) \ln \left( 
\frac{-q^{2}}{\lambda ^{2}}\right) M_{1}  \label{M2}
\end{equation}
and 
\begin{equation}
M_{3}=\frac{\alpha Z}{\pi }\frac{\epsilon _{3}}{|{\bf {\text{p}}_{3}|}}\ln
\left( \frac{\epsilon _{3}+|{\bf {\text{p}}_{3}|}}{m}\right) \ln \left( 
\frac{-q^{2}}{\lambda ^{2}}\right) M_{1}  \label{M3}
\end{equation}

By contrast, in \cite{tsai}, in addition to the approximation just
described, a soft-photon approximation is made in the infrared denominators:
Specifically, when $k=0$ the factor $(k-q)^{2}-\lambda ^{2}$ is set equal to 
$q^{2}-\lambda ^{2}$ and when $k=q$ the factor $k^{2}-\lambda ^{2}$ is set
equal to $q^{2}-\lambda ^{2}$, thus giving two terms and reducing the
four-point function to three-point functions: 
\begin{equation}
M_{2}=-\frac{\alpha Z}{2\pi }\left[ K(p_{2},-p_{1})+K(p_{4},-p_{3})\right]
M_{1}  \label{M2ts}
\end{equation}
and 
\begin{equation}
M_{3}=\frac{\alpha Z}{2\pi }\left[ K(p_{2},p_{3})+K(p_{4},p_{1})\right] M_{1}
\label{M3ts}
\end{equation}
(Note \cite{tsai}, eqs.(II.9) and (II.11)). The infrared divergent terms
(those with a factor $\ln $ $\lambda ^{2}$) are, for $M_{2}$, the same in (%
\ref{M2}) and (\ref{M2ts}), and, for $M_{3}$, the same in (\ref{M3}) and (%
\ref{M3ts}) . However, (\ref{M2ts}) and (\ref{M3ts}) differ significantly
from (\ref{M2}) and (\ref{M3}). These latter expressions are functions of
the momentum transfer, $q^{2}$. The integrals $K(p_{i},p_{j})$, on the other
hand, are functions only of the scalar invariants $p_{i}^{2},$ $p_{j}^{2}$
and $p_{i}\cdot p_{j}$. In (3.21) and (3.22), $M_{2}$ and $M_{3}$ therefore
depend only on the initial and final electron energies, and not on the
momentum transfer ( $p_{2}\cdot p_{1}=$ $p_{4}\cdot p_{3}=\epsilon _{1}M;$ $%
p_{2}\cdot p_{3}=$ $p_{4}\cdot p_{1}=\epsilon _{3}M$).

From (\ref{vp2})-(\ref{vp3}), (\ref{M6bar}), (\ref{M6bar0}), (\ref{M5bar}), (%
\ref{M2}) and (\ref{M3}), we can now write the square of the matrix element
for elastic scattering, including the radiative correction to order $\alpha $%
. Assuming $-q^{2}\gg m^{2},$ and including only electron-positron pairs in
the vacuum polarization matrix element, we have 
\begin{eqnarray}
|{\cal M}|^{2} &=&|M_{1}|^{2}\left\{ 
\begin{array}{c}
1+\frac{\alpha }{\pi }\left[ {\frac{13}{6}}\ln \left( \frac{-q^{2}}{m^{2}}%
\right) -{\frac{28}{9}}-K(p_{1},p_{3})+K(p_{1},p_{1})\right] \\ 
-\frac{2\alpha Z}{\pi }\ln \eta \ln \left( \frac{-q^{2}}{\lambda ^{2}}\right)
\\ 
+\frac{\alpha Z^{2}}{\pi }\left[ -K(p_{2},p_{4})+K(p_{2},p_{2})\right]
\end{array}
\right\}  \label{M2elas} \\
&&+2\text{Re}\left\{ M_{1}^{\dagger }\overline{M}_{6}^{(1)}\right\} 
\nonumber
\end{eqnarray}

\subsection{Contribution of proton form factor}

Finally, we consider the contribution of the term $2$Re$\left\{
M_{1}^{\dagger }\overline{M}_{6}^{(1)}\right\} $, coming from the inclusion
of form factors for the proton and integration over the entire range of
four-momenta of the virtual photon in the proton vertex correction.
Equations (\ref{gpsggps}) and (\ref{gpssgps}) define the functions $%
G_{1}(q^{2}),G_{2}(q^{2}),X_{1}(q^{2}),$ and $X_{2}(q^{2}).$ From (\ref{M0M1}%
) we may write $M_{6}^{(1)}=M_{6}-M_{6}^{(0)}$, {\it i.e.}, the term $%
M_{6}^{(1)}$ is obtained from the full proton vertex correction by
subtracting the infrared divergent matrix element $M_{6}^{(0)}$ which is
independent of the proton form factor. We therefore define $G_{1}^{^{\prime
}}(q^{2})$ and $X_{2}^{^{\prime }}(q^{2})$ to be the expressions $%
G_{1}(q^{2})$ and $X_{2}(q^{2})$ from which we have omitted the terms with
factor $\phi _{1}(\lambda ^{2})$ We then write, from (\ref{gpsggps}) and (%
\ref{gpssgps}), 
\begin{equation}
\overline{(g+s)g(g+s)}\,^{\prime }=F(q^{2})\left[ \left( G_{1}^{^{\prime
}}(q^{2})-G_{1}^{^{\prime }}(0)\right) \gamma _{\mu }+G_{2}(q^{2})\frac{%
i\sigma _{\mu \nu }q^{\nu }}{2M}\right]
\end{equation}
and 
\begin{equation}
\overline{(g+s)s(g+s)}\,^{\prime }=\kappa F(q^{2})\left[ X_{1}(q^{2})\gamma
_{\mu }+\left( X_{2}^{^{\prime }}(q^{2})-G_{1}^{^{\prime }}(0)\right) \frac{%
i\sigma _{\mu \nu }q^{\nu }}{2M}\right]
\end{equation}
We now define the functions $F_{1g}(q^{2}),F_{2g}(q^{2}),F_{1s}(q^{2}),$ and 
$F_{2s}(q^{2})$ by 
\begin{equation}
F(q^{2})\left[ \left( G_{1}^{^{\prime }}(q^{2})-G_{1}^{^{\prime }}(0)\right)
\gamma _{\mu }+G_{2}(q^{2})\frac{i\sigma _{\mu \nu }q^{\nu }}{2M}\right]
\equiv F_{1g}(q^{2})\gamma _{\mu }+\kappa F_{2g}(q^{2})\frac{i\sigma _{\mu
\nu }q^{\nu }}{2M}
\end{equation}
\begin{equation}
\kappa F(q^{2})\left[ X_{1}(q^{2})\gamma _{\mu }+\left( X_{2}^{^{\prime
}}(q^{2}\}-G_{1}^{^{\prime }}(0)\right) \frac{i\sigma _{\mu \nu }q^{\nu }}{2M%
}\right] \equiv F_{1s}(q^{2})\gamma _{\mu }+\kappa F_{2s}(q^{2})\frac{%
i\sigma _{\mu \nu }q^{\nu }}{2M}
\end{equation}
Then with the further definitions 
\begin{equation}
\widetilde{F}_{1}(q^{2})\equiv F_{1g}(q^{2})+F_{1s}(q^{2})
\end{equation}
\begin{equation}
\widetilde{F}_{2}(q^{2})\equiv F_{2g}(q^{2})+F_{2s}(q^{2})
\end{equation}
\begin{equation}
\widetilde{\Gamma }_{\mu }\equiv \widetilde{F}_{1}(q^{2})\gamma _{\mu
}+\kappa \widetilde{F}_{2}(q^{2})\frac{i\sigma _{\mu \nu }q^{\nu }}{2M}
\end{equation}
we have (apart from factors) 
\begin{equation}
\overline{M}_{6}^{(1)}=\frac{\alpha Z^{2}}{2\pi }\left\langle p_{3}|\gamma
^{\mu }|p_{1}\right\rangle \left\langle p_{4}|\widetilde{\Gamma }_{\mu
}|p_{2}\right\rangle
\end{equation}
and 
\begin{equation}
2\text{Re}\left\{ M_{1}^{\dagger }\overline{M}_{6}^{(1)}\right\} =\frac{%
\alpha Z^{2}}{\pi }\left( \left\langle p_{3}|\gamma ^{\nu
}|p_{1}\right\rangle \left\langle p_{4}|\Gamma _{\nu }|p_{2}\right\rangle
\right) ^{\dagger }\left( \left\langle p_{3}|\gamma ^{\mu
}|p_{1}\right\rangle \left\langle p_{4}|\widetilde{\Gamma }_{\mu
}|p_{2}\right\rangle \right)
\end{equation}
This has the same form as 
\begin{equation}
M_{1}^{\dagger }M_{1}=\left( \left\langle p_{3}|\gamma ^{\nu
}|p_{1}\right\rangle \left\langle p_{4}|\Gamma _{\nu }|p_{2}\right\rangle
\right) ^{\dagger }\left( \left\langle p_{3}|\gamma ^{\mu
}|p_{1}\right\rangle \left\langle p_{4}|\Gamma _{\mu }|p_{2}\right\rangle
\right)
\end{equation}
with the exception of the replacement $\Gamma _{\mu }\rightarrow \widetilde{%
\Gamma }_{\mu }$ in the right-hand term. Thus, in place of the Rosenbluth
cross section, (\ref{rosen}), obtained from $\sum_{spins}M_{1}^{\dagger
}M_{1}$, we have 
\begin{equation}
\sum_{spins}2\text{Re}\left\{ M_{1}^{\dagger }\overline{M}_{6}^{(1)}\right\}
=\frac{\alpha ^{2}\cos ^{2}\frac{\theta }{2}}{4\epsilon _{1}^{2}\eta \sin
^{4}\left( \theta /2\right) }\left( \frac{\alpha Z^{2}}{\pi }\right) \left\{ 
\begin{array}{l}
\bigskip
\end{array}
\right\}
\end{equation}
where 
\begin{equation}
\left\{ 
\begin{array}{l}
\bigskip
\end{array}
\right\} =\left( F_{1}\widetilde{F}_{1}-\frac{\kappa ^{2}q^{2}}{4M^{2}}F_{2}%
\widetilde{F}_{2}\right) -\frac{q^{2}}{2M^{2}}\left( F_{1}+\kappa
F_{2}\right) \left( \widetilde{F}_{1}+\kappa \widetilde{F}_{2}\right) \tan
^{2}\frac{\theta }{2}
\end{equation}
The purely elastic cross section, including radiative corrections to order $%
\alpha ,$ can thus be written as 
\begin{equation}
\left( \frac{d\sigma _{0}}{d\Omega }\right) \left( 1+\delta
_{el}^{(0)}+\delta _{el}^{(1)}\right)
\end{equation}
where 
\begin{eqnarray}
\delta _{el}^{(0)} &=&\frac{\alpha }{\pi }\left[ -\left[ \ln \left( \frac{%
-q^{2}}{m^{2}}\right) -1\right] \ln \left( \frac{m^{2}}{\lambda ^{2}}\right)
+{\frac{13}{6}}\ln \left( \frac{-q^{2}}{m^{2}}\right) -{\frac{28}{9}}-\frac{1%
}{2}\ln ^{2}\left( \frac{-q^{2}}{m^{2}}\right) {+}\frac{\pi ^{2}}{6}\right] 
\nonumber \\
&-&\frac{2\alpha Z}{\pi }\ln \eta \ln \left( \frac{-q^{2}}{\lambda ^{2}}%
\right)  \label{del0} \\
&+&\frac{\alpha Z^{2}}{\pi }\left[ -\left( \frac{\epsilon _{4}}{|{\bf {\text{%
p}}_{4}|}}\ln x-1\right) \ln \left( \frac{M^{2}}{\lambda ^{2}}\right) +\frac{%
\epsilon _{4}}{|{\bf {\text{p}}_{4}|}}\left[ -\ln x\ln \left( \frac{\rho ^{2}%
}{M^{2}}\right) +\frac{1}{2}\ln ^{2}x+2L(-\frac{1}{x})+\frac{\pi ^{2}}{6}%
\right] \right]  \nonumber
\end{eqnarray}
and 
\begin{equation}
\delta _{el}^{(1)}=\frac{\alpha Z^{2}}{\pi }\left\{ \frac{\left( F_{1}%
\widetilde{F}_{1}-\frac{\kappa ^{2}q^{2}}{4M^{2}}F_{2}\widetilde{F}%
_{2}\right) -\frac{q^{2}}{2M^{2}}\left( F_{1}+\kappa F_{2}\right) \left( 
\widetilde{F}_{1}+\kappa \widetilde{F}_{2}\right) \tan ^{2}\frac{\theta }{2}%
}{\left[ \left( F_{1}^{2}-\frac{\kappa ^{2}q^{2}}{4M^{2}}F_{2}^{2}\right) -%
\frac{q^{2}}{2M^{2}}\left( F_{1}+\kappa F_{2}\right) ^{2}\tan ^{2}\frac{%
\theta }{2}\right] }\right\}  \label{del1}
\end{equation}

\section{Inelastic Cross Section}

In this section we calculate the inelastic cross section, i.e., the
contribution of soft photon emission by the initial and final electron and
proton to the radiative correction. The relevant diagrams, with
corresponding matrix elements $M_{b1}$ and $M_{b2}$, are shown in Fig.\ \ref
{fig2}. These matrix elements are given by

\begin{eqnarray}
M_{b1} &=&-iZe^{3}(2\pi )^{4}\delta ^{4}(p_{3}+p_{4}+k-p_{1}-p_{2})\frac{mM}{%
\sqrt{2\omega \epsilon _{1}\epsilon _{3}\epsilon _{2}\epsilon _{4}}} 
\nonumber \\
&&\times \overline{u}(p_{3})\left[ \epsilon\!\!\!/\frac{1}{%
p\!\!\!/_{3}+k\!\!\!/-m+i\epsilon }\gamma _{\mu }+\gamma _{\mu }\frac{1}{%
p\!\!\!/_{1}-k\!\!\!/-m+i\epsilon }\epsilon\!\!\!/\right] u(p_{1})
\label{4.1} \\
&&\times \overline{u}(p_{4})\Gamma ^{\mu }u(p_{2})\frac{1}{%
(p_{1}-p_{3}-k)^{2}+i\epsilon }  \nonumber
\end{eqnarray}
\begin{eqnarray}
M_{b2} &=&iZ^{2}e^{3}(2\pi )^{4}\delta ^{4}(p_{3}+p_{4}+k-p_{1}-p_{2})\frac{%
mM}{\sqrt{2\omega \epsilon _{1}\epsilon _{3}\epsilon _{2}\epsilon _{4}}}%
\overline{u}(p_{3})\gamma _{\mu }u(p_{1})  \nonumber \\
&&\,\times \overline{u}(p_{4})\left[ \epsilon\!\!\!/\frac{1}{%
p\!\!\!/_{4}+k\!\!\!/-m+i\epsilon }\Gamma ^{\mu }+\Gamma ^{\mu }\frac{1}{%
p\!\!\!/_{2}-k\!\!\!/-m+i\epsilon }\epsilon\!\!\!/\right] u(p_{2})\frac{1}{%
(p_{1}-p_{3})^{2}+i\epsilon }  \label{4.2}
\end{eqnarray}

Making the soft photon approximation, we rationalize the denominators and
drop terms of relative order $k$ in the numerator and denominator (but not
in the delta function), giving

\begin{eqnarray}
M_{b1}+M_{b2} &=&-iZe^{3}(2\pi )^{4}\delta ^{4}(p_{3}+p_{4}+k-p_{1}-p_{2}) 
\frac{mM}{\sqrt{2\omega \epsilon _{1}\epsilon _{3}\epsilon _{2}\epsilon _{4}}%
} \frac{1}{q^{2}}  \nonumber \\
&&\,\times \overline{u}(p_{3})\gamma _{\mu }u(p_{1})\,\overline{u}%
(p_{4})\Gamma ^{\mu }u(p_{2})\left( \frac{p_{3}\cdot \epsilon }{p_{3}\cdot k}%
-\frac{p_{1}\cdot \epsilon }{p_{1}\cdot k}-Z\frac{p_{4}\cdot \epsilon }{%
p_{4}\cdot k}+Z\frac{p_{2}\cdot \epsilon }{p_{2}\cdot k}\right)  \label{4.3}
\end{eqnarray}

\thinspace \thinspace \thinspace \thinspace \thinspace \thinspace \thinspace
\thinspace \thinspace \thinspace \thinspace \thinspace \thinspace \thinspace
\thinspace \thinspace \thinspace \thinspace \thinspace

\thinspace \thinspace The cross section for soft bremsstrahlung then follows
by squaring the matrix element $M_{b1}+M_{b2}$ , dividing by the incident
flux and the transition rate and multiplying by the number of final states.
Summing over photon polarizations, we then have

\begin{eqnarray}
d\sigma _{b} &=&-\frac{Z^{2}e^{6}}{(2\pi )^{9}}\frac{m^{2}M^{2}}{\sqrt{%
(p_{1}\cdot p_{2})^{2}-m^{2}M^{2}}}\sum_{spins}\int \frac{d^{3}p_{3}}{%
\epsilon _{3}}\frac{d^{3}p_{4}}{\epsilon _{4}}\frac{d^{3}k}{2\omega }(2\pi
)^{4}\delta ^{4}(p_{3}+p_{4}+k-p_{1}-p_{2})  \nonumber \\
&&\times \left| \overline{u}(p_{3})\gamma _{\mu }u(p_{1})\,\overline{u}%
(p_{4})\Gamma ^{\mu }u(p_{2})\right| ^{2}\frac{1}{(q^{2})^{2}}\left( \frac{%
p_{3}}{p_{3}\cdot k}-\frac{p_{1}}{p_{1}\cdot k}-Z\frac{p_{4}}{p_{4}\cdot k}+Z%
\frac{p_{2}}{p_{2}\cdot k}\right) ^{2}  \label{4.4}
\end{eqnarray}

The range of integration in the above expression is determined by the
experimental conditions. We assume, as in \cite{tsai}, that the final proton
and emitted photon are undetected; the range of integration in energy and
angle of the final electron is determined by the entrance slit and
spectrometer. We integrate first over $d^{3}p_{4}$ and are then left with a
single delta function relating the variables of ${\bf k}$ and ${\bf p}_{3}$:
Writing 
\begin{equation}
\frac{d^{3}p_{4}}{2\epsilon _{4}}=\int_{0}^{\infty }d\epsilon _{4}\,\delta
(p_{4}^{2}-M^{2})\,d^{3}p_{4}=\int d^{4}p_{4}\,\delta
(p_{4}^{2}-M^{2})\,\theta (p_{4}^{0})  \label{4.5}
\end{equation}
with 
\begin{equation}
t\equiv p_{1}+p_{2}-p_{3}=p_{4}+k  \label{4.7}
\end{equation}
we then have 
\begin{eqnarray}
d\sigma _{b} &=&-\frac{Z^{2}e^{6}}{(2\pi )^{5}}\frac{m^{2}M^{2}}{\sqrt{%
(p_{1}\cdot p_{2})^{2}-m^{2}M^{2}}}\sum_{spins}\int \frac{d^{3}p_{3}}{%
\epsilon _{3}}\int \frac{d^{3}k}{\omega }\,\delta \left(
(t-k)^{2}-M^{2}\right) \theta (\epsilon _{4})  \nonumber \\
&&\times \left| \overline{u}(p_{3})\gamma _{\mu }u(p_{1})\,\overline{u}%
(p_{4})\Gamma ^{\mu }u(p_{2})\right| ^{2}\frac{1}{(q^{2})^{2}}\left( \frac{%
p_{3}}{p_{3}\cdot k}-\frac{p_{1}}{p_{1}\cdot k}-Z\frac{p_{4}}{p_{4}\cdot k}+Z%
\frac{p_{2}}{p_{2}\cdot k}\right) ^{2}  \label{4.8}
\end{eqnarray}
in which $p_{4}=p_{1}+p_{2}-p_{3}-k$. We may then transform to the special
frame S$^{0}$ (defined by {\bf t} $=0$), in which the delta function in (\ref
{4.8}) is independent of the angle at which the photon is emitted. There 
\begin{equation}
(t-k)^{2}-M^{2}=t_{0}^{2}-2t_{0}\omega +\lambda
^{2}-M^{2}=0;\,\,\,\,\,\,\,\,\,\,\,\,\,\,\,\,\,\,\,\,\,\,\,t_{0}=\epsilon
_{1}+\epsilon _{2}-\epsilon _{3}  \label{4.9}
\end{equation}
The photon energy is then given solely by the final electron energy. The
procedure used in \cite{tsai} is to integrate next over the photon energy
and angle in S$^{0}$ and then transform back to the lab frame to integrate
over the energy and angle of the final electron. Instead, we remain in the
special frame and integrate first over $\epsilon _{3}$, the delta function
giving $\epsilon _{3}$ in terms of $\omega $. However, the range of photon
energies is assumed to be sufficiently small compared to all other energies
that we can set $\epsilon _{3}$ equal to its value for elastic scattering
throughout the integrand. In addition, we take the angular range of the
final electron to be sufficiently small that we can take some average value
for these angles and neglect any variation of the integrand over this
angular range. Similarly, we neglect $k$ in the above expression for $p_{4}$%
. We may then take $\left| \overline{u}(p_{3})\gamma _{\mu }u(p_{1})\,%
\overline{u}(p_{4})\Gamma ^{\mu }u(p_{2})\right| ^{2}\diagup (q^{2})^{2}$
outside of the integration, giving 
\begin{eqnarray}
d\sigma _{b} &=&-\frac{Z^{2}e^{6}}{(2\pi )^{5}}\frac{m^{2}M^{2}}{\sqrt{%
(p_{1}\cdot p_{2})^{2}-m^{2}M^{2}}}\sum_{spins}\left| \overline{u}%
(p_{3})\gamma _{\mu }u(p_{1})\,\overline{u}(p_{4})\Gamma ^{\mu
}u(p_{2})\right| ^{2}\frac{1}{(q^{2})^{2}}  \nonumber \\
&&\times \frac{\left| \text{{\bf p}}_{3}\right| }{2M}\int^{^{\prime }}\frac{%
d^{3}k}{\omega }\,\left( \frac{p_{3}}{p_{3}\cdot k}-\frac{p_{1}}{p_{1}\cdot k%
}-Z\frac{p_{4}}{p_{4}\cdot k}+Z\frac{p_{2}}{p_{2}\cdot k}\right) ^{2}
\label{4.10}
\end{eqnarray}
The term $2M$ in the denominator in (\ref{4.10}) comes from the delta
function in (\ref{4.8}), which contributes the factor $|d\{\delta \left(
(t-k)^{2}-M^{2}\right) \}/d\epsilon _{3}|=2|t_{0}-\omega |$ from (\ref{4.9}%
). Again neglecting terms of order $k,$ we note from (\ref{4.7}) that in S$%
^{0}$, $t_{0}-\omega =\epsilon _{4}=M$. Comparing (\ref{4.4}), (\ref{xs})
and (\ref{ope}) and noting that in arriving at (\ref{4.10}) we have
neglected terms of order $k$ in $p_{3}$ and $p_{4}$, we have 
\begin{equation}
d\sigma _{b}=-\frac{\alpha }{4\pi ^{2}}\,d\sigma _{0}\int^{^{\prime }}\frac{%
d^{3}k}{\omega }\,\left( \frac{p_{3}}{p_{3}\cdot k}-\frac{p_{1}}{p_{1}\cdot k%
}-Z\frac{p_{4}}{p_{4}\cdot k}+Z\frac{p_{2}}{p_{2}\cdot k}\right) ^{2}
\label{4.11}
\end{equation}
where $\omega =\sqrt{{\bf k}^{2}+\lambda ^{2}}$. There then remains the
integration over photon energy (restricted to $\left| {\bf k}\right| \leq
\Delta \epsilon $) and angles. The relevant integrals have been evaluated by
't Hooft and Veltman \cite{thooft}, Sec. 7. We give here only their final
result, rewritten using our metric; the essential steps in the derivation
are given in their work. They define 
\begin{equation}
L_{ij}=\int^{^{\prime }}\frac{d^{3}k}{\omega }\frac{1}{(p_{i}\cdot
k)(p_{j}\cdot k)}  \label{4.12}
\end{equation}
in terms of which 
\begin{equation}
d\sigma _{b}=-\frac{\alpha }{4\pi ^{2}}\,d\sigma _{0}\left\{ 
\begin{array}{c}
m^{2}L_{11}+m^{2}L_{33}-2p_{1}\cdot p_{3}L_{13} \\ 
+Z\left( -2p_{1}\cdot p_{2}L_{12}+2p_{3}\cdot p_{2}L_{32}+2p_{1}\cdot
p_{4}L_{14}-2p_{3}\cdot p_{4}L_{34}\right) \\ 
+Z^{2}\left( M^{2}L_{22}+M^{2}L_{44}-2p_{2}\cdot p_{4}L_{24}\right)
\end{array}
\right\}  \label{4.12a}
\end{equation}
As shown in \cite{thooft}, Sec.7, for the case in which the momenta $p_{i}$
and $p_{j}$ are all on the mass shell, the integrals $L_{ij}$ can, provided $%
p_{i}$ is not a multiple $p_{j}$, be written in the form 
\begin{equation}
L_{ij}=\frac{2\pi }{\sqrt{(p_{i}\cdot p_{j})^{2}-m_{i}^{2}m_{j}^{2}}}\left\{
S_{ij}^{(1)}+S_{ij}^{(2)}\right\}  \label{4.13}
\end{equation}
where 
\begin{equation}
S_{ij}^{(1)}=2\,\ln \left( \frac{p_{i}\cdot p_{j}+\sqrt{(p_{i}\cdot
p_{j})^{2}-m_{i}^{2}m_{j}^{2}}}{m_{i}m_{j}}\right) \ln \left( \frac{2\Delta
\epsilon }{\lambda }\right)  \label{4.14}
\end{equation}
and 
\begin{eqnarray}
S_{ij}^{(2)} &=&\ln ^{2}\left( \frac{\beta _{i}}{m_{i}\sqrt{t^{2}}}\right)
-\ln ^{2}\left( \frac{\beta _{j}}{m_{j}\sqrt{t^{2}}}\right)  \nonumber \\
&&+L\left( 1-\frac{\beta _{i}\,l\cdot t}{t^{2}\gamma _{ij}}\right) +L\left(
1-\frac{\,m_{i}^{2}l\cdot t}{\beta _{i}\gamma _{ij}}\right)  \nonumber \\
&&-L\left( 1-\frac{\beta _{j}\,l\cdot t}{\alpha t^{2}\gamma _{ij}}\right)
-L\left( 1-\frac{m_{j}^{2}\,l\cdot t}{\alpha \beta _{j}\gamma _{ij}}\right)
\label{4.15}
\end{eqnarray}
in which 
\begin{equation}
\alpha =\frac{p_{i}\cdot p_{j}+\sqrt{(p_{i}\cdot
p_{j})^{2}-m_{i}^{2}m_{j}^{2}}}{m_{i}^{2}}\text{; \thinspace \thinspace
\thinspace \thinspace \thinspace \thinspace \thinspace \thinspace \thinspace
\thinspace \thinspace \thinspace \thinspace \thinspace \thinspace \thinspace
\thinspace }\,\,\,l=\alpha p_{i}-p_{j}  \label{4.16}
\end{equation}
\begin{equation}
\beta _{i,j}\equiv p_{i,j}\cdot t+\sqrt{(p_{i,j}\cdot t)^{2}-m_{i,j}^{2}t^{2}%
}\text{; \thinspace \thinspace \thinspace \thinspace \thinspace \thinspace
\thinspace \thinspace \thinspace \thinspace \thinspace \thinspace \thinspace
\thinspace \thinspace \thinspace \thinspace }\gamma _{ij}\equiv \sqrt{%
(p_{i}\cdot p_{j})^{2}-m_{i}^{2}m_{j}^{2}}  \label{4.17}
\end{equation}
The evaluation of (\ref{4.12}) for $p_{i}=p_{j}$ is straightforward. The
result written in terms of relativistic invariants is 
\begin{equation}
L_{ii}=\frac{4\pi }{m_{i}^{2}}\left[ \ln \left( \frac{2\Delta \epsilon }{%
\lambda }\right) -\frac{p_{i}\cdot t}{\sqrt{(p_{i}\cdot t)^{2}-m_{i}^{2}t^{2}%
}}\ln \left( \frac{\beta _{i}}{m_{i}\sqrt{t^{2}}}\right) \right]
\label{4.18}
\end{equation}
In \cite{thooft}, Sec. 7, the expression for $S_{ij}^{(2)}$ is evaluated in
the frame S$^{0}$, defined by {\bf t}$=0$. Since we want finally to express
the cross section in terms of lab frame energies and momenta, we have, in (%
\ref{4.15}), written $S_{ij}^{(2)}$ in terms of relativistic invariants.

The terms of leading order in $\ln \lambda $ are apparent in (\ref{4.14})
and (\ref{4.18}). Substituting these in (\ref{4.12a}) gives the infrared
divergent terms in $d\sigma _{b}$. They are cancelled exactly by the $\ln
\lambda $ terms in the elastic cross section.

We next express $d\sigma _{b}$ in terms of lab frame energies. To that end,
we assume that $\Delta \epsilon $ is less than any of the other energies and
therefore now neglect $k$ in (\ref{4.7}), taking $p_{4}$ to be given by its
value for elastic (non-radiative) scattering: 
\begin{equation}
t=p_{1}+p_{2}-p_{3}=p_{4}  \label{4.19}
\end{equation}
(Note that for $\Delta \epsilon \rightarrow 0,$ $S_{ij}^{(2)}$ remains
finite; the only singularity is confined to the term $\ln (2\Delta \epsilon
/\lambda )$, evident in $S_{ij}^{(1)}$.) The relativistic invariants in $%
L_{ij}$ can then be written simply in terms of lab frame energies: 
\begin{eqnarray}
p_{1}\cdot p_{4} &=&p_{2}\cdot p_{3}=M\epsilon _{3}  \nonumber \\
p_{3}\cdot p_{4} &=&p_{2}\cdot p_{1}=M\epsilon _{1}  \label{4.20} \\
p_{1}\cdot p_{3} &=&-\frac{1}{2}q^{2}+m^{2}  \nonumber \\
p_{2}\cdot p_{4} &=&-\frac{1}{2}q^{2}+M^{2}  \nonumber
\end{eqnarray}
Further, we express $\Delta \epsilon $, the maximum momentum of the photon
in the frame $S^{0}$, in terms of the final electron detector acceptance in
the lab frame, $\Delta E$: 
\begin{equation}
\Delta \epsilon =\eta \Delta E  \label{4.21}
\end{equation}
Details of the derivation are given in Appendix E. Substituting (\ref{4.19})
and (\ref{4.20}) in (\ref{4.14}), (\ref{4.15}) and (\ref{4.18}) then gives $%
L_{ij}$ in terms of lab frame energies and momenta. We note that although $%
L_{ij}$ as defined in (\ref{4.12}) is clearly symmetric in $i$ and $j$, the
expression for $S_{ij}^{(2)}$ in (\ref{4.15}) is not manifestly symmetric;
the form of the expression for $L_{ij}$ is rather different from that for $%
L_{ji}$. In writing the explicit expressions for the terms in $L_{ij}$, we
choose $i$ and $j$ such that $L_{ij}$ simplifies readily for lab frame
electron energies and momentum transfers which are very large compared to
the electron rest mass. When $m_{i}\neq m_{j}$, this is achieved by choosing 
$i$ and $j$ such that $m_{i}=m$ and $m_{j}=M$ (note below that we have
written $S_{32}^{(2)}$ ).

At this point we make the high energy approximation as defined in section
II, in which case the above expressions for $S_{ij}^{(1)}$ and $S_{ij}^{(2)}$
simplify considerably. For $S_{ij}^{(1)}$ this is straightforward. We give
the results for $S_{ij}^{(2)}$ in Appendix F. Substituting these in $L_{ij}$%
, the high energy approximation for the inelastic cross section given in (%
\ref{4.12a}) then becomes 
\begin{equation}
d\sigma _{b}=\frac{\alpha }{\pi }\,d\sigma _{0}\left\{ 
\begin{array}{c}
\left[ \ln \left( \frac{-q^{2}}{m^{2}}\right) -1\right] \ln \left( \frac{%
(2\eta \Delta E)^{2}}{\lambda ^{2}}\right) \\ 
-\left[ \ln \left( \frac{-q^{2}}{m^{2}}\right) -1\right] \ln \left( \frac{%
4\epsilon _{1}\epsilon _{3}}{m^{2}}\right) \\ 
+\frac{1}{2}\ln ^{2}\left( \frac{-q^{2}}{m^{2}}\right) -\frac{1}{2}\ln
^{2}\eta +L(\cos ^{2}\frac{1}{2}\theta )-\frac{1}{3}\pi ^{2} \\ 
+2Z\left[ 
\begin{array}{c}
\ln \eta \,\ln \left( \frac{(2\eta \Delta E)^{2}}{\lambda ^{2}}\right) -\ln
\eta \,\ln x \\ 
+L(1-\frac{\eta }{x})-L(1-\frac{1}{\eta x})
\end{array}
\right] \\ 
+Z^{2}\left[ 
\begin{array}{c}
\left( \frac{\epsilon _{4}}{|{\bf \text{p}}_{4}|}\ln x-1\right) \ln \left( 
\frac{(2\eta \Delta E)^{2}}{\lambda ^{2}}\right) \\ 
-\frac{\epsilon _{4}}{|{\bf \text{p}}_{4}|}\left[ \ln ^{2}x-\ln x+L(1-\frac{1%
}{x^{2}})\right] -1
\end{array}
\right]
\end{array}
\right\}  \label{4.22}
\end{equation}

\section{Radiative Corrections to elastic electron-proton scattering}

The results given in (\ref{del0}), (\ref{del1}), and (\ref{4.22}) may be
added to give the radiative correction, $\delta $. The analytic expression
is given below in (\ref{dsig}) and (\ref{delta}). Numerical evaluation of
the radiative correction for various values of the pertinent parameters
(initial beam energy, momentum transfer, final electron detector resolution,
and target nucleus) are given in Tables \ref{table1}, \ref{table2}, and \ref
{table3}. We note that the infrared ($\ln \lambda $) terms, which appear in
both the purely elastic ( \ref{del0}) and inelastic (\ref{4.22})
contributions to the radiative correction, cancel exactly when added to give
the cross section for elastic electron-proton scattering with radiative
corrections to first order in $\alpha $: 
\begin{equation}
d\sigma =d\sigma _{0}(1+\delta )  \label{dsig}
\end{equation}
where 
\begin{eqnarray}
\delta &=&\frac{\alpha }{\pi }\left[ \frac{13}{6}\ln \left( \frac{-q^{2}}{%
m^{2}}\right) -\frac{28}{9}-\left[ \ln \left( \frac{-q^{2}}{m^{2}}\right)
-1\right] \ln \left( \frac{4\epsilon _{1}\epsilon _{3}}{(2\eta \Delta E)^{2}}%
\right) -\frac{1}{2}\ln ^{2}\eta +L(\cos ^{2}\frac{1}{2}\theta )-\frac{\pi
^{2}}{6}\right]  \nonumber \\
&&+\frac{2\alpha Z}{\pi }\left[ -\ln \eta \,\ln \left( \frac{-q^{2}x}{(2\eta
\Delta E)^{2}}\right) +L\left( 1-\frac{\eta }{x}\right) -L\left( 1-\frac{1}{%
\eta x}\right) \right]  \label{delta} \\
&&+\frac{\alpha Z^{2}}{\pi }\left[ 
\begin{array}{c}
\frac{\epsilon _{4}}{|{\bf \text{p}}_{4}|}\left( -\frac{1}{2}\ln ^{2}x-\ln
x\ln \left( \frac{\rho ^{2}}{M^{2}}\right) +\ln x\right) -\left( \frac{%
\epsilon _{4}}{|{\bf \text{p}}_{4}|}\ln x-1\right) \ln \left( \frac{M^{2}}{%
(2\eta \Delta E)^{2}}\right) +1 \\ 
+\frac{\epsilon _{4}}{|{\bf \text{p}}_{4}|}\left( -L\left( 1-\frac{1}{x^{2}}%
\right) +2L\left( -\frac{1}{x}\right) +\frac{\pi ^{2}}{6}\right)
\end{array}
\right]  \nonumber \\
&&+\delta _{el}^{(1)}  \nonumber
\end{eqnarray}
Here, $\delta _{el}^{(1)}$ is the contribution coming from the inclusion of
form factors for the proton and integration over the entire range of
four-momenta of the virtual photon in the proton vertex correction (see (\ref
{M0M1}), (\ref{M6bar}), (\ref{M6bar0}), (\ref{M2elas})); it is thus not
included in the analysis given in \cite{tsai} and \cite{mo}, denoted here as
the soft photon approximation. Moreover, $\delta _{el}^{(1)}$ has no
infrared divergent terms; these are all included in the soft photon
approximation.

In Tables \ref{table1}, \ref{table2}, and \ref{table3} we compare the values
of the radiative correction, $\delta $, calculated in this paper (denoted by
MTj) with those given by Mo and Tsai in \cite{mo} for various kinematics.
The initial beam energies and momentum transfers have been chosen to
correspond to experiments proposed or already performed at Jefferson Lab 
\cite{jlab} and SLAC \cite{SLAC}. The final electron detector acceptance, $%
\Delta E$, has been taken throughout to be one percent of the final electron
energy, $\epsilon _{3}$. In the form factors (see(\ref{ff})), the parameter $%
\Lambda $ has been chosen to be 700 MeV/$c$ throughout. The contribution of
the terms in (\ref{delta}) are grouped according to the power of $Z$ which
appears there as a factor. The numerical value of each of these groups of
terms is given in the rows denoted by $Z^{0},Z^{1},Z^{2}$. Values given in
the column MTj in the row $Z^{2}$ do not include the contribution of the
proton form factor (which are contained in $\delta _{el}^{(1)})$; they are
given for comparison with the values in \cite{mo}. In the range of energies
and momentum transfers considered here, the correction $\delta _{el}^{(1)}$,
due to the finite size of the nucleon (and integration over the entire range
of four-momenta of the virtual photon in the proton vertex correction), is
found in general to be much smaller than the other contributions with factor 
$Z^{2}$, labeled explicitly in (\ref{delta}) and in Tables \ref{table1}, \ref
{table2}, and \ref{table3}. The values given in these tables include only
the contribution of electron-positron pairs in the vacuum polarization; the
contribution of muon and tau pairs is given by (\ref{vp2}) and (\ref{vp3}).

The curves in Figs.\ \ref{fig3} and \ref{fig4} illustrate the two aspects of
the present work: (1) the contribution of nucleonic size effects to the
radiative correction, and (2) the improvement of the mathematical treatment
of the integrations given in the work of Mo and Tsai \cite{tsai}, \cite{mo}.
The nucleonic size effects are all contained in the term $\delta _{el}^{(1)}$%
, (Eq.(\ref{del1})); its contribution relative to the overall radiative
correction factor, $(1+\delta _{\text{MTj}})$, is given by the dashed curves
marked VTX, where VTX = 100$\times \delta _{el}^{(1)}/(1+\delta _{\text{MTj}%
})$. The dotted curve shows D0 = 100$\times ($ $\delta _{\text{MTj}%
}^{(0)}-\delta _{\text{Tsai}})/(1+\delta _{\text{MTj}})$, which is that part
of the difference between the radiative correction given by Mo and Tsai \cite
{mo} and the one given in this paper due solely to the improvement of the
mathematical treatment of the integrations. Here, $\delta _{\text{MTj}}^{(0)}
$ is the radiative correction given in (\ref{delta}), {\it excluding} the
term $\delta _{el}^{(1)}$. It will be noticed that VTX is always positive,
and that for most of the range of allowed momentum transfers, D0 is
negative. Thus their sum, which is the difference between the radiative
correction $\delta _{\text{MTj}}$ given in this paper in (\ref{delta}), and $%
\delta _{\text{Tsai}}$, given in \cite{mo}, is rather small except for the
region corresponding to large scattering angles. This sum is given by the
solid curves marked D, where D = D0 + VTX = 100$\times ($ $\delta _{\text{MTj%
}}-\delta _{\text{Tsai}})/(1+\delta _{\text{MTj}})$.

\smallskip

\section{Conclusion}

We have calculated the radiative correction to elastic electron-proton
scattering to lowest order in $\alpha $ using a hadronic model which
includes the finite size of the nucleon. The contribution from the emission
of real soft photons by the electron and the proton is calculated exactly.
The contributions of the box and crossed-box (two-photon exchange) diagrams
are calculated in a soft photon approximation which is less drastic than
that employed in \cite{tsai}. A number of observations may be made from the
values given in Tables \ref{table1}, \ref{table2}, and \ref{table3}. First,
the contributions of the electron vertex correction, vacuum polarization,
and real soft photon emission by the electron (the terms in (\ref{delta})
with factor $\alpha /\pi $) dominate the radiative correction $\delta $.
Since our expression for these terms differs from that given by Mo and Tsai 
\cite{mo} solely in that they have omitted the term $\left( \alpha /\pi
\right) \left[ L(\cos ^{2}\frac{1}{2}\theta )-\frac{\pi ^{2}}{6}\right] $ in
(\ref{delta}), we find values for $\delta $ which differ from theirs by at
most 2\% for the initial energies and momentum transfers considered here
(note that $-\frac{\pi ^{2}}{6}\leq L(\cos ^{2}\frac{1}{2}\theta )-\frac{\pi
^{2}}{6}\leq 0$). Further, we note that, except for the proton and at the
higher energies considered here, the contribution of $\delta _{el}^{(1)}$ is
negligible. However, for the two highest energies, $\delta _{el}^{(1)}$ is
between 2\% and 3\% of the factor ($1+\delta $) by which the uncorrected
cross section must be multiplied, and hence should be considered in
precision measurements for electron-proton scattering at energies above 8
GeV. As an empirical guide, we find that $\delta _{el}^{(1)}=0.02(1+\delta )$
for initial energies and scattering angles satisfying $\epsilon _{1}\sin
\theta \approx 8$ for beam energies between 8 and 16 GeV. Finally, we note
that a considerable simplification of the expression in (\ref{delta}) occurs
if, in addition to the last two terms multiplying $\alpha /\pi $, we neglect
the last two terms multiplying $2\alpha Z/\pi $ as well as the last three
terms multiplying $\alpha Z^{2}/\pi $, each of these sets of terms being
always less than $\pi ^{2}/6$ in magnitude. From this study we see that at
the energies and momentum transfers considered here, the nucleonic finite
size effects are rather small but are expected to become more important at
higher energies. The corrections due to the improvement of the high energy
behavior of the radiative corrections as described in this paper are not
negligible and need to be taken into account at the energies and momentum
transfers we have considered.

\begin{acknowledgments}
It is a pleasure to acknowledge the valued assistance of W.C. Parke in the
realization of this paper.
\end{acknowledgments}

\appendix

\section{Elastic scattering amplitudes to order $\alpha ^{2}$}

Using the notation given in section II, the matrix elements corresponding to
the various one-loop diagrams shown in Fig.\ \ref{fig1} are

\begin{equation}
M_{5}=Ze^{2}\overline{u}(p_{3})\Lambda _{\mu }(p_{3},p_{1})u(p_{1})\frac{(-i)%
}{q^{2}+i\epsilon }\overline{u}(p_{4})\Gamma ^{\mu }(q^{2})u(p_{2})
\label{A3}
\end{equation}
\begin{equation}
M_{6}=Z^{3}e^{2}\overline{u}(p_{3})\gamma _{\mu }u(p_{1})\frac{(-i)}{%
q^{2}+i\epsilon }\overline{u}(p_{4})\Lambda ^{\mu }(p_{4},p_{2})u(p_{2})
\label{A4}
\end{equation}
where 
\begin{equation}
\Lambda _{\mu }(p_{3},p_{1})=-ie^{2}\int \frac{d^{4}k}{(2\pi )^{4}}\frac{1}{%
k^{2}-\lambda ^{2}+i\epsilon }\gamma ^{\nu }\frac{1}{(p\!\!\!/_{3}-k\!\!%
\!/-m+i\epsilon )}\gamma _{\mu }\frac{1}{(p\!\!\!/_{1}-k\!\!\!/-m+i\epsilon )%
}\gamma _{\nu }  \label{A5}
\end{equation}
\newline
\newline
\begin{eqnarray}
\Lambda ^{\mu }(p_{4},p_{2}) &=&-ie^{2}\int \frac{d^{4}k}{(2\pi )^{4}}\frac{1%
}{k^{2}-\lambda ^{2}+i\epsilon }\Gamma ^{\nu }(k^{2})\frac{1}{%
(p\!\!\!/_{4}-k\!\!\!/-M+i\epsilon )}\Gamma ^{\mu }(q^{2})\allowbreak 
\nonumber \\[0.11in]
&&\,\,\,\,\,\,\,\,\,\,\,\,\,\,\,\,\,\,\,\,\,\,\,\,\,\,\,\,\,\,\,\,\,\,\,\,%
\times \frac{1}{(p\!\!\!/_{2}-k\!\!\!/-M+i\epsilon )}\Gamma _{\nu }(k^{2})
\label{A6}
\end{eqnarray}
\begin{eqnarray}
M_{2} &=&(Ze^{2})^{2}\int \frac{d^{4}k}{(2\pi )^{4}}\,\frac{1}{k^{2}-\lambda
^{2}+i\epsilon }\,\,\frac{1}{(k+q)^{2}-\lambda ^{2}+i\epsilon }  \nonumber \\
&&\,\,\,\,\,\,\,\,\,\,\,\,\,\,\,\,\,\,\,\,\,\times \left[ \overline{u}%
(p_{3})\gamma _{\nu }\frac{1}{p\!\!\!/_{1}-k\!\!\!/-m+i\epsilon }\gamma
_{\mu }u(p_{1})\right]  \label{box} \\
&&\,\,\,\,\,\,\,\,\,\,\,\,\,\,\,\,\,\,\,\,\times \,\left[ \overline{u}%
(p_{4})\Gamma ^{\nu }((k+q)^{2})\frac{1}{p\!\!\!/_{2}+k\!\!\!/-M+i\epsilon }%
\Gamma ^{\mu }(k^{2})u(p_{2})\right]  \nonumber
\end{eqnarray}
\begin{eqnarray}
M_{3} &=&(Ze^{2})^{2}\int \frac{d^{4}k}{(2\pi )^{4}}\,\frac{1}{k^{2}-\lambda
^{2}+i\epsilon }\,\,\frac{1}{(k+q)^{2}-\lambda ^{2}+i\epsilon }  \nonumber \\
&&\,\,\,\,\,\,\,\,\,\,\,\,\,\,\,\,\,\,\,\,\,\times \left[ \overline{u}%
(p_{3})\gamma _{\nu }\frac{1}{p\!\!\!/_{1}-k\!\!\!/-m+i\epsilon }\gamma
_{\mu }u(p_{1})\right]  \label{cbx} \\
&&\,\,\,\,\,\,\,\,\,\,\,\,\,\,\,\,\,\,\,\,\times \,\left[ \overline{u}%
(p_{4})\Gamma ^{\mu }(k^{2})\frac{1}{p\!\!\!/_{4}-k\!\!\!/-M+i\epsilon }%
\Gamma ^{\nu }((k+q)^{2})u(p_{2})\right]  \nonumber
\end{eqnarray}
The matrix element, $M_{4}$, for vacuum polarization is, after charge
renormalization, related simply to the matrix element $M_{1}$ given above: 
\begin{equation}
M_{4}=\Pi (q^{2})M_{1}  \label{vp}
\end{equation}
For a fermion loop in the photon propagator, $\Pi (q^{2})$ is given in \cite
{bj} in terms of an integral which can be evaluated in closed form \cite
{tsai2}, giving 
\begin{equation}
\Pi (q^{2})=\Pi ^{f}(q^{2}/m_{i}^{2})=\frac{\alpha }{3\pi }\left\{ \left( 1-%
\frac{u}{2}\right) \sqrt{1+u}\log \left( \frac{\sqrt{1+u}+1}{\sqrt{1+u}-1}%
\right) +u-\frac{5}{3}\right\}  \label{vp2}
\end{equation}
in which $m_{i}$ is the mass of the fermion and $u=4m_{i}^{2}/(-q^{2}).$

For $-q^{2}/m_{i}^{2}\gg 1$ this gives 
\begin{equation}
\Pi ^{f}(q^{2}/m_{i}^{2})=\frac{\alpha }{\pi }\left\{ \frac{1}{3}\ln \left( 
\frac{-q^{2}}{m_{i}^{2}}\right) -\frac{5}{9}\right\}  \label{vp2a}
\end{equation}
If we include the vacuum polarization amplitudes from particle-antiparticle
loops of different masses, as has been done in several experimental analyses 
\cite{SLAC}, then we have 
\begin{equation}
M_{4}=M_{1}\sum_{i}\Pi ^{f}(q^{2}/m_{i}^{2})  \label{vp3}
\end{equation}
In principle, once one includes particle-antiparticle pairs of mass greater
than the electron mass in the vacuum polarization amplitudes, one should
consider bosons as well as fermions. The matrix elements for vacuum
polarization for a pair of structureless spin zero bosons in the closed
loop, first given by Feynman \cite{feynman}, may be found in a more
accessible form in a paper of Tsai \cite{tsai2}. The result, corresponding
to the equation above for fermions, is 
\begin{equation}
M_{4}^{boson}=M_{1}\sum_{i}\Pi ^{b}(q^{2}/m_{i}^{2})  \label{vpboson}
\end{equation}
where 
\begin{equation}
\Pi ^{b}(q^{2}/m_{i}^{2})=\frac{\alpha }{3\pi }\left\{ \frac{1}{2}\sqrt{1+u}%
\log \left( \frac{\sqrt{1+u}+1}{\sqrt{1+u}-1}\right) -u-\frac{4}{3}\right\}
\label{vp4}
\end{equation}
\medskip

A more complete discussion of vacuum polarization should include a
consideration of pion structure as well as the contribution of spin-one
bosons, in particular the $\rho $ meson. A detailed discussion of the
hadronic contribution to vacuum polarization may be found in connection with
calculations of the anomalous magnetic moment of the muon \cite{kino} and in
connection with radiative corrections to high energy electron-positron
collider experiments \cite{fleischer}.

\smallskip

\section{Proton vertex correction}

As noted in Sec. IIIA, the terms $ggg,gsg,...$ can be expressed in terms of
the integrals given in (3.1). There, the integrals $I_{0},J_{0},$ and $K_{0}$
are scalars and hence are functions of the scalars $p_{2}^{2},\,p_{4}^{2},$%
and $p_{2}\cdot p_{4}$ (and, of course, $\lambda ^{2}$ and $\Lambda ^{2})$.
Since we have on-shell particles in the initial and final states $%
(p_{2}^{2}=p_{4}^{2}=M^{2})$, these integrals are functions of $M^{2}$ and $%
q^{2}$. The integrals $I_{\mu }$ and $J_{\mu }$ are vectors and hence in
principal can be written in the form 
\begin{equation}
I_{\mu }=ap_{2\mu }+bp_{4\mu }  \label{B3}
\end{equation}
with a similar equation for $J_{\mu }$ , where $a$ and $b$ are functions of $%
M^{2}$ and $q^{2}$. However, the calculation is simplified greatly if we
express $I_{\mu }$ in terms of the four-vectors $\rho =p_{4}+p_{2}$ (which
is symmetric in $p_{4}$ and $p_{2}$) and $q=p_{4}-p_{2}$ (which is
antisymmetric in $p_{4}$ and $p_{2}$), {\it i.e}. $I_{\mu }=A\rho _{\mu
}+Bq_{\mu }$. Here $A$ and $B$ are functions of $M^{2}$ and $q^{2}$ and
hence are symmetric in $p_{4}$ and $p_{2}$. Further, since the integrands
for the vectors $I_{\mu }$ and $J_{\mu }$ are symmetric in $p_{4}$ and $%
p_{2} $, it follows that $B=0.$ We thus have 
\begin{equation}
I_{\mu }=A\rho _{\mu }  \label{B4}
\end{equation}
and a similar equation for $J_{\mu }$. These same considerations of symmetry
allow for the simplification of the tensors $I_{\mu \nu }$ and $J_{\mu \nu }$%
, which are also symmetric functions of $p_{4}$ and $p_{2}$. We can
therefore write 
\begin{equation}
I_{\mu \nu }=a_{1}\rho _{\mu }\rho _{\nu }+a_{2}q_{\mu }q_{\nu }+a_{3}g_{\mu
\nu }
\end{equation}
and a similar equation for $J_{\mu \nu }$. That the terms $\rho _{\mu
}q_{\nu }$ and $q_{\mu }\rho _{\nu }$ are absent follows directly by
multiplying $I_{\mu \nu }$ successively by $\rho ^{\mu }q^{\nu }$ and $%
q^{\mu }\rho ^{\nu }$, using $\rho \cdot q=0$ and the fact that $I_{\mu \nu
}\rho ^{\mu }q^{\nu }$ and $I_{\mu \nu }q^{\mu }\rho ^{\nu }$ are
antisymmetric in $p_{2}$ and $p_{4}$. Multiplying $I_{\mu }$ ($J_{\mu }$) by 
$\rho ^{\mu }$, and $I_{\mu \nu }$ ($J_{\mu \nu }$) successively by $\rho
^{\mu }\rho ^{\nu },q^{\mu }q^{\nu },$ and $g^{\mu \nu }$, the coefficients
in the expressions for $I_{\mu },J_{\mu },I_{\mu \nu }$ and $J_{\mu \nu }$
may be expressed in terms of their moments, defined by 
\begin{equation}
g_{1}=\frac{1}{\rho ^{2}}I_{\mu }\rho ^{\mu
},\,\,\,\,\,\,\,\,\,\,\,\,\,\,\,\,\,\,\,\,\,\,\,\,\,\,\,\,\,\,\,h_{1}=\frac{1%
}{\rho ^{2}}J_{\mu }\rho ^{\mu }  \label{g1h1}
\end{equation}
\begin{equation}
g_{11}=\frac{1}{\rho ^{4}}I_{\mu \nu }\rho ^{\mu }\rho ^{\nu
},\,\,\,\,\,\,\,\,\,\,\,\,\,\,\,\,\,\,\,\,\,\,\,\,\,\,\,\,\,\,\,h_{11}=\frac{%
1}{\rho ^{4}}J_{\mu \nu }\rho ^{\mu }\rho ^{\nu }
\end{equation}
\begin{equation}
g_{22}=\frac{1}{\rho ^{4}}I_{\mu \nu }q^{\mu }q^{\nu
},\,\,\,\,\,\,\,\,\,\,\,\,\,\,\,\,\,\,\,\,\,\,\,\,\,\,\,\,\,\,\,h_{22}=\frac{%
1}{\rho ^{4}}J_{\mu \nu }q^{\mu }q^{\nu }  \label{g22h22}
\end{equation}

Conversely, the integrals in (\ref{3.1}) may be written in terms of the
moments:

\begin{equation}
I_{\mu }=\rho _{\mu
}g_{1},\,\,\,\,\,\,\,\,\,\,\,\,\,\,\,\,\,\,\,\,\,\,\,\,\,\,\,\,\,\,\,\,J_{%
\mu }=\rho _{\mu }h_{1}\,  \label{B10}
\end{equation}
\begin{eqnarray}
I_{\mu \nu } &=&\rho _{\mu }\rho _{\nu }\left[ g_{11}-\frac{1}{2\rho ^{2}}%
(h_{0}-\rho ^{2}g_{11}-q^{2}g_{22})\right]  \nonumber  \\
&&+q_{\mu }q_{\nu }\left[ g_{22}-\frac{1}{2q^{2}}(h_{0}-\rho
^{2}g_{11}-q^{2}g_{22})\right] \\
&&+g_{\mu \nu }\left[ \frac{1}{2}(h_{0}-\rho ^{2}g_{11}-q^{2}g_{22})\right]
\nonumber
\end{eqnarray}
\begin{eqnarray}
J_{\mu \nu } &=&\rho _{\mu }\rho _{\nu }\left[ h_{11}-\frac{1}{2\rho ^{2}}%
(k_{0}-\rho ^{2}h_{11}-q^{2}h_{22})\right]  \nonumber \\
&&+q_{\mu }q_{\nu }\left[ h_{22}-\frac{1}{2q^{2}}(k_{0}-\rho
^{2}h_{11}-q^{2}h_{22})\right] \\
&&+g_{\mu \nu }\left[ \frac{1}{2}(k_{0}-\rho ^{2}h_{11}-q^{2}h_{22})\right] 
\nonumber
\end{eqnarray}
where, for convenience of notation, we have defined 
\begin{equation}
g_{0}=I_{0},\,\,\,\,\,\,\,\,\,\,\,\,\,\,h_{0}=J_{0},\,\,\,\,\,\,\,\,\,\,\,\,%
\,k_{0}=K_{0}  \label{B13}
\end{equation}
The terms $ggg,gsg,...$ can be expressed in terms of these moments.
Substituting (\ref{cur}) and (\ref{ff}) in the expression for the proton
vertex correction, Eq. (\ref{A6}), and substituting this in turn in the
matrix element $M_{6}$, Eq. (\ref{A4}), the integrals are all of the form
given in Eq. (\ref{3.1}), which are in turn expressed in terms of the
moments by Eqs. (\ref{B10})-(\ref{B13}). We make use of the fact that in $%
M_{6}$ the vertex correction is taken between free spinors, and finally
express $p_{2}$ and $p_{4}$ in terms of $\rho $ and $q$. We find 
\begin{equation}
ggg=-ie^{2}F(q^{2})\left\{ 
\begin{array}{c}
\left[ 
\begin{array}{c}
2(2M^{2}-q^{2})g_{0}-4(2M^{2}-q^{2})g_{1} \\ 
-2(8M^{2}+q^{2})g_{11}-2(q^{4}/\rho ^{2})g_{22}+8(M^{2}/\rho ^{2})h_{0}
\end{array}
\right] \gamma _{\mu } \\ 
+\left[ -8M^{2}g_{1}+24M^{2}g_{11}+8M^{2}(q^{2}/\rho
^{2})g_{22}-8(M^{2}/\rho ^{2})h_{0}\right] \frac{i\sigma _{\mu \nu }q^{\nu }%
}{2M}
\end{array}
\right\}
\end{equation}
\begin{equation}
gsg=-ie^{2}\kappa F(q^{2})\left\{ 
\begin{array}{c}
\left[ -2q^{2}g_{1}\right] \gamma _{\mu } \\ 
+\left[ 2(2M^{2}-q^{2})g_{0}-4(2M^{2}-q^{2})g_{1}\right] \frac{i\sigma _{\mu
\nu }q^{\nu }}{2M}
\end{array}
\right\}
\end{equation}
\begin{equation}
ggs+sgg=-2ie^{2}\kappa F(q^{2})\left\{ 
\begin{array}{c}
\left[ -12M^{2}g_{11}-(q^{4}/\rho ^{2})g_{22}+2(1+2M^{2}/\rho
^{2})h_{0}-3h_{1}\right] \gamma _{\mu } \\ 
+\left[ 
\begin{array}{c}
(16M^{2}-q^{2})g_{11}+(q^{2}/\rho ^{2})(8M^{2}-q^{2})g_{22} \\ 
-4(1+M^{2}/\rho ^{2})h_{0}+3h_{1}
\end{array}
\right] \frac{i\sigma _{\mu \nu }q^{\nu }}{2M}
\end{array}
\right\}
\end{equation}
\begin{equation}
gss+ssg=-2ie^{2}\kappa ^{2}F(q^{2})\left\{ 
\begin{array}{c}
(q^{2}/4M^{2})\left[ 
\begin{array}{c}
-(8M^{2}+q^{2})g_{11}-(q^{4}/\rho ^{2})g_{22} \\ 
+(-2+4M^{2}/\rho ^{2})h_{0}+h_{1}
\end{array}
\right] \gamma _{\mu } \\ 
+\left[ 3q^{2}g_{11}+4M^{2}(q^{2}/\rho ^{2})g_{22}-(q^{2}/\rho
^{2})h_{0}-h_{1}\right] \frac{i\sigma _{\mu \nu }q^{\nu }}{2M}
\end{array}
\right\}
\end{equation}
\begin{equation}
sgs=-ie^{2}\kappa ^{2}F(q^{2})\left\{ 
\begin{array}{c}
\left[ 
\begin{array}{c}
-(8M^{2}+q^{2})g_{11}-(q^{4}/\rho ^{2})g_{22}+(2+q^{2}/\rho ^{2})h_{0} \\ 
-(8M^{2}+q^{2})h_{11}/4M^{2}-(q^{4}/\rho ^{2})h_{22}/4M^{2}+(-1+q^{2}/\rho
^{2})k_{0}/4M^{2}
\end{array}
\right] \gamma _{\mu } \\ 
+\left[ 
\begin{array}{c}
12M^{2}g_{11}+4M^{2}(q^{2}/\rho ^{2})g_{22}-2(1+2M^{2}/\rho ^{2})h_{0} \\ 
+3h_{11}+4(q^{2}/\rho ^{2})h_{22}-k_{0}/\rho ^{2}
\end{array}
\right] \frac{i\sigma _{\mu \nu }q^{\nu }}{2M}
\end{array}
\right\}
\end{equation}
\begin{equation}
sss=-ie^{2}\kappa ^{3}F(q^{2})\left\{ 
\begin{array}{c}
(q^{2}/4M^{2})\left[ 
\begin{array}{c}
-12M^{2}g_{11}-4M^{2}(q^{2}/\rho ^{2})g_{22}+2(1+2M^{2}/\rho ^{2})h_{0} \\ 
-4h_{1}+3h_{11}+(q^{2}/\rho ^{2})h_{22}-k_{0}/\rho ^{2}
\end{array}
\right] \gamma _{\mu } \\ 
+\left[ 
\begin{array}{c}
2(2M^{2}+q^{2})g_{11}+4M^{2}(q^{2}/\rho ^{2})g_{22} \\ 
-(4M^{2}/\rho ^{2}+q^{2}/2M^{2})h_{0}+(q^{2}/M^{2})h_{1} \\ 
-2(2M^{2}+q^{2})h_{11}/4M^{2}+2(q^{2}/\rho ^{2})(2M^{2}-q^{2})h_{22}/4M^{2}
\\ 
+(q^{2}/\rho ^{2})k_{0}/4M^{2}
\end{array}
\right] {\frac{i\sigma _{\mu \nu }q^{\nu }}{2M}}
\end{array}
\right\}
\end{equation}
The expressions do not depend on the particular form of the form factors; we
have assumed only that $F_{1}=F_{2}=F$. However, for form factors of the
form given in Eq. (\ref{ff}), the moments may all be expressed more simply
in terms of the functions $C(\Lambda ^{2}):$%
\begin{equation}
\left\{ C_{0}(\Lambda ^{2});C_{\mu }(\Lambda ^{2});C_{\mu \nu }(\Lambda
^{2})\right\} =\int d^{4}k\,\left\{ 1;k_{\mu };k_{\mu \nu }\right\}
/D(\Lambda ^{2})  \label{B20}
\end{equation}
Using the identity 
\begin{equation}
\frac{1}{k^{2}-\lambda ^{2}}\left( \frac{-\Lambda ^{2}}{k^{2}-\Lambda ^{2}}%
\right) ^{m}=\frac{(-\Lambda ^{2})^{m}}{(m-1)!}T^{m-1}\left\{ \frac{1}{%
\Lambda ^{2}-\lambda ^{2}}\left[ \frac{1}{k^{2}-\Lambda ^{2}}-\frac{1}{%
k^{2}-\lambda ^{2}}\right] \right\}
\end{equation}
with $T\equiv \frac{\partial }{\partial (\Lambda ^{2})}$, we can write 
\begin{equation}
\left\{ I\right\} =N_{m}^{^{\prime }}(\Lambda ^{2})^{m}T^{m-1}\left\{ \frac{%
C(\Lambda ^{2})-C(\lambda ^{2})}{\Lambda ^{2}-\lambda ^{2}}\right\} ,
\label{B23}
\end{equation}
where 
\begin{equation}
N_{m}^{^{\prime }}=\frac{(-1)^{m}}{(m-1)!(2\pi )^{4}}.
\end{equation}
In Eq. (\ref{B23}) $I$ and $C$ denote any one of $I_{0},I_{\mu },I_{\mu \nu
} $ and $C_{0},C_{\mu },C_{\mu \nu }$ respectively. We see from Eq. (\ref
{B23}) that terms in $C(\Lambda ^{2})$ which are independent of $\Lambda
^{2} $ do not appear in the expression for $I$ . In particular, this applies
to $C_{\mu \nu }(\Lambda ^{2})$, which may be evaluated using either
dimensional regularization or a convergence factor. The infinities in $%
C_{\mu \nu }(\Lambda ^{2})$ are indeed independent of $\Lambda ^{2}$, thus
giving a finite result for $I_{\mu \nu }$ as it should. In similar fashion,
we have 
\begin{equation}
\left\{ J\right\} =N_{m}^{^{\prime }}(\Lambda ^{2})^{m}T^{m-1}\left\{ \frac{%
\Lambda ^{2}C(\Lambda ^{2})-\lambda ^{2}C(\lambda ^{2})}{\Lambda
^{2}-\lambda ^{2}}\right\}  \label{B26}
\end{equation}
in which $J$ and $C$ denote any one of $J_{0},J_{\mu },J_{\mu \nu }$ and $%
C_{0},C_{\mu },C_{\mu \nu }$ respectively. We see from Eq. (\ref{B26}) that
any terms in $C(\Lambda ^{2})$ which are independent of $\Lambda ^{2}$ do
not appear in the expression for $J$ provided $m>1$. Finally, for $K_{0}$ we
get 
\begin{equation}
K_{0}=N_{m}^{^{\prime }}(\Lambda ^{2})^{m}T^{m-1}\left\{ \frac{\Lambda
^{4}C_{0}(\Lambda ^{2})-\lambda ^{4}C_{0}(\lambda ^{2})}{\Lambda
^{2}-\lambda ^{2}}\right\}  \label{B28}
\end{equation}

We note that, apart from trivial factors, the integrals in (\ref{B20}) are
the three-point functions defined in \cite{thooft}, Eq.(5.1), and \cite
{passarino}, Eq.(E.1); $C_{0}$ has been evaluated in terms of Spence
functions in \cite{thooft}. The details of the algebra in \cite{thooft} and 
\cite{passarino} being rather lengthy, we choose instead to evaluate the
integrals in (\ref{B20}) using Feynman parameters, writing 
\begin{equation}
\frac{1}{D(\Lambda ^{2})}=2\int_{0}^{1}\int_{0}^{1}\frac{x\,dx\,dy}{%
[k^{2}-2xk\cdot p_{y}-\Lambda ^{2}(1-x)+i\epsilon ]^{3}},  \label{B29}
\end{equation}
where $p_{y}=p_{2}y+p_{4}(1-y)$. Substituting Eq. (\ref{B29}) in Eq. (\ref
{B20}) and shifting the integration variable $k\,\,(k-xp_{y}\rightarrow k)$
we then have

\begin{eqnarray}
&&\left\{ C_{0}(\Lambda ^{2});C_{\mu }(\Lambda ^{2});C_{\mu \nu }(\Lambda
^{2})\right\}  \nonumber \\
&=&-i\pi ^{2}\int_{0}^{1}\int_{0}^{1}\,dx\,dy\frac{\left\{ x;x^{2}p_{y\mu
};x^{3}p_{y\mu }p_{y\nu }+\frac{1}{2}g_{\mu \nu }\lambda ^{2}(x-\frac{1}{2}%
x^{2})\right\} }{x^{2}p_{y}^{2}+\Lambda ^{2}(1-x)}\,\,+\,\chi _{\mu \nu },
\label{B32}
\end{eqnarray}
where, throughout this section, we denote by $\,\chi _{\mu \nu }$ any terms
which are independent of $\Lambda $ . Rewriting $p_{y}=\frac{1}{2}\rho +%
\frac{1}{2}q(1-2y)$ in Eqs. (\ref{B29}) and (\ref{B32}) we may, neglecting
terms which are independent of $\Lambda $, express $C_{0},C_{\mu }$ and $%
C_{\mu \nu }$ in terms of the functions 
\begin{equation}
\phi _{k}(\Lambda ^{2})\equiv \int_{0}^{1}\int_{0}^{1}\,\frac{x^{k}\,dx\,dy}{%
p_{y}^{2}x^{2}+\Lambda ^{2}(1-x)}.  \label{phik}
\end{equation}
We get 
\begin{equation}
C_{0}=-i\pi ^{2}\phi _{1}(\lambda ^{2});\,\,\,C_{\mu }=-i\pi ^{2}\frac{1}{2}%
\rho _{\mu }\phi _{2}(\lambda ^{2})  \label{C0l}
\end{equation}
\begin{eqnarray}
C_{\mu \nu } &=&-i\pi ^{2}\left[ \frac{1}{4}\rho _{\mu }\rho _{\nu }\phi
_{3}(\lambda ^{2})-\frac{1}{4}\frac{\rho ^{2}}{q^{2}}q_{\mu }q_{\nu }\phi
_{3}(\lambda ^{2})+\right.  \nonumber \\
&&\left. -\frac{q_{\mu }q_{\nu }}{q^{2}}\lambda ^{2}[\phi _{1}(\lambda
^{2})-\phi _{2}(\lambda ^{2})]+\frac{1}{2}g_{\mu \nu }\lambda ^{2}[\phi
_{1}(\lambda ^{2})-\frac{1}{2}\phi _{2}(\lambda ^{2})]\right] .
\end{eqnarray}
As shown in Appendix D, the functions $\phi _{k}$ obey a three-term
inhomogeneous recursion, which is used to calculate $\phi _{k}$ for $k>1$: 
\begin{eqnarray}
(k+1)\rho ^{2}\phi _{k+2}(\Lambda ^{2})-2(2k+1)\Lambda ^{2}\phi
_{k+1}(\Lambda ^{2})+ &&4k\Lambda ^{2}\phi _{k}(\Lambda ^{2})  \label{recur}
\\
=\frac{2\rho }{\rho _{1}}\ln \left( \frac{\rho +\rho _{1}}{\rho -\rho _{1}}%
\right) &&+2\Lambda ^{2}\left[ \phi _{k+1}^{(0)}(\Lambda ^{2})-2\phi
_{k}^{(0)}(\Lambda ^{2})\right]  \nonumber
\end{eqnarray}
Here

\begin{equation}
\phi _{k}^{(0)}(\Lambda ^{2})\equiv \phi _{k}(\Lambda
^{2})|_{-q^{2}=0}=\int_{0}^{1}\frac{x^{k}\,dx}{M^{2}x^{2}+(1-x)\Lambda ^{2}}
\label{B28p}
\end{equation}
The functions $\phi _{k}^{(0)}(\Lambda ^{2})$ may in turn be calculated from
the recursion 
\begin{equation}
M^{2}\phi _{k+2}^{(0)}(\Lambda ^{2})-\Lambda ^{2}\phi _{k+1}^{(0)}(\Lambda
^{2})+\Lambda ^{2}\phi _{k}^{(0)}(\Lambda ^{2})=\frac{1}{k+1}  \label{recur0}
\end{equation}
To implement the recursions (\ref{recur}) and (\ref{recur0}) we need 
\begin{equation}
\phi _{0}^{(0)}(\Lambda ^{2})=\frac{1}{\Lambda \Lambda _{1}}\ln \left( \frac{%
\Lambda +\Lambda _{1}}{\Lambda -\Lambda _{1}}\right)
\end{equation}
and 
\begin{equation}
\phi _{1}^{(0)}(\Lambda ^{2})=\frac{1}{2M^{2}}\left[ \ln \frac{M^{2}}{%
\Lambda ^{2}}+\frac{\Lambda }{\Lambda _{1}}\ln \left( \frac{\Lambda +\Lambda
_{1}}{\Lambda -\Lambda _{1}}\right) \right]  \label{phi10Lam}
\end{equation}
which follow from Eq. (\ref{B28p}), and $\phi _{1}(\Lambda ^{2})$, which can
be expressed in terms of dilogarithms (Spence functions)(see Appendix D): 
\begin{equation}
\phi _{1}(\Lambda ^{2})=\frac{1}{\rho \rho _{1}}\left\{ L\left( 1-\frac{1}{xy%
}\right) -L\left( 1-\frac{x}{y}\right) -2\ln (x)\ln (1+\frac{1}{y})\right\}
\label{phi1}
\end{equation}
where 
\begin{equation}
L(z)=-\int_{0}^{z}\frac{\ln (1-t)}{t}\,dt
\end{equation}
\begin{equation}
x=\frac{\rho +\rho _{1}}{\rho -\rho _{1}}=\frac{(\rho +\rho _{1})^{2}}{4M^{2}%
}
\end{equation}
\begin{equation}
y=\frac{\Lambda +\Lambda _{1}}{\Lambda -\Lambda _{1}}=\frac{(\Lambda
+\Lambda _{1})^{2}}{4M^{2}}
\end{equation}
We will also want to take the limit $\lambda \rightarrow 0$. Neglecting all
terms which vanish in this limit, we find, 
\begin{equation}
\phi _{1}(\lambda ^{2})\,\,_{\overrightarrow{\lambda \rightarrow 0}}\,\,%
\frac{1}{\rho \rho _{1}}\left\{ -2L\left( -\frac{1}{x}\right) -\frac{\pi ^{2}%
}{6}-\frac{1}{2}\ln ^{2}x+\ln x\,\ln \left( \frac{\rho ^{2}}{\lambda ^{2}}%
\right) \right\}  \label{phi1lam}
\end{equation}
\begin{equation}
\phi _{1}^{(0)}(\lambda ^{2})\,\,_{\overrightarrow{\lambda \rightarrow 0}}%
\frac{1}{M^{2}}\ln \left( \frac{M}{\lambda }\right)  \label{phi10lam}
\end{equation}
and for $k>1,$%
\begin{equation}
\phi _{k}(0)=\frac{2}{(k-1)\rho \rho _{1}}\ln x  \label{B36}
\end{equation}
\begin{equation}
\phi _{k}^{(0)}(0)=\frac{1}{(k-1)M^{2}}
\end{equation}
We then find 
\begin{equation}
g_{0}=-N_{1}\phi _{1}(\lambda ^{2})+N_{m}(\Lambda ^{2})^{m}T^{m-1}\left\{ 
\frac{1}{\Lambda ^{2}}\phi _{1}(\Lambda ^{2})\right\}
\end{equation}
\begin{equation}
g_{1}=\frac{1}{2}N_{m}(\Lambda ^{2})^{m}T^{m-1}\left\{ \frac{1}{\Lambda ^{2}}%
\left[ \phi _{2}(\Lambda ^{2})-\phi _{2}(0)\right] \right\}
\end{equation}
\begin{eqnarray}
g_{11} &=&\frac{1}{4}N_{m}(\Lambda ^{2})^{m}T^{m-1}\left\{ \frac{1}{\Lambda
^{2}}\left[ \phi _{3}(\Lambda ^{2})-\phi _{3}(0)\right] \right\} \\
&&+\frac{1}{2\rho ^{2}}N_{m}(\Lambda ^{2})^{m}T^{m-1}\left\{ \phi
_{1}(\Lambda ^{2})-\frac{1}{2}\phi _{2}(\Lambda ^{2})\right\}  \nonumber
\end{eqnarray}
\begin{eqnarray}
g_{22} &=&-\frac{\rho ^{2}}{4q^{2}}N_{m}(\Lambda ^{2})^{m}T^{m-1}\left\{ 
\frac{1}{\Lambda ^{2}}\left[ \phi _{3}(\Lambda ^{2})-\phi _{3}(0)\right]
\right\} \\
&&-\frac{1}{2q^{2}}N_{m}(\Lambda ^{2})^{m}T^{m-1}\left\{ \phi _{1}(\Lambda
^{2})-\frac{3}{2}\phi _{2}(\Lambda ^{2})\right\}  \nonumber
\end{eqnarray}
\begin{equation}
h_{0}=N_{m}(\Lambda ^{2})^{m}T^{m-1}\left\{ \phi _{1}(\Lambda ^{2})\right\}
\end{equation}
\begin{equation}
h_{1}=\frac{1}{2}N_{m}(\Lambda ^{2})^{m}T^{m-1}\left\{ \phi _{2}(\Lambda
^{2})\right\}
\end{equation}
\begin{eqnarray}
h_{11} &=&\frac{1}{4}N_{m}(\Lambda ^{2})^{m}T^{m-1}\left\{ \phi _{3}(\Lambda
^{2})\right\} \\
&&+\frac{1}{2\rho ^{2}}N_{m}(\Lambda ^{2})^{m}T^{m-1}\left\{ \Lambda
^{2}\left[ \phi _{1}(\Lambda ^{2})-\frac{1}{2}\phi _{2}(\Lambda ^{2})\right]
\right\}  \nonumber
\end{eqnarray}
\begin{eqnarray}
h_{22} &=&-\frac{\rho ^{2}}{4q^{2}}N_{m}(\Lambda ^{2})^{m}T^{m-1}\left\{
\phi _{3}(\Lambda ^{2})\right\} \\
&&-\frac{1}{2q^{2}}N_{m}(\Lambda ^{2})^{m}T^{m-1}\left\{ \Lambda ^{2}\left[
\phi _{1}(\Lambda ^{2})-\frac{3}{2}\phi _{2}(\Lambda ^{2})\right] \right\} 
\nonumber
\end{eqnarray}
\begin{equation}
k_{0}=N_{m}(\Lambda ^{2})^{m}T^{m-1}\left\{ \Lambda ^{2}\phi _{1}(\Lambda
^{2})\right\}
\end{equation}
where 
\begin{equation}
N_{m}=-i\pi ^{2}N_{m}^{^{\prime }}
\end{equation}
The terms $ggg,gsg,...sss$ can now be expressed more simply in terms of the
functions $\phi _{k}$. We get 
\begin{eqnarray}
ggg &=&-ie^{2}F(q^{2})\left\{ 
\begin{array}{c}
\begin{array}{c}
-2(2M^{2}-q^{2})N_{1}\phi _{1}(\lambda ^{2}) \\ 
+2(2M^{2}-q^{2})\left[ S^{m-1}\left\{ \frac{1}{\Lambda ^{2}}\phi
_{1}(\Lambda ^{2})\right\} -S^{m-1}\left\{ \frac{1}{\Lambda ^{2}}\left[ \phi
_{2}(\Lambda ^{2})-\phi _{2}(0)\right] \right\} \right]
\end{array}
\\ 
+S^{m-1}\left\{ \phi _{2}(\Lambda ^{2})\right\} -4M^{2}S^{m-1}\left\{ \frac{1%
}{\Lambda ^{2}}\left[ \phi _{3}(\Lambda ^{2})-\phi _{3}(0)\right] \right\}
\end{array}
\right\} \gamma _{\mu }  \label{ggg2} \\
&&-ie^{2}F(q^{2})\left\{ -4M^{2}S^{m-1}\left\{ \frac{1}{\Lambda ^{2}}\left[
\phi _{2}(\Lambda ^{2})-\phi _{2}(0)\right] \right\} +4M^{2}S^{m-1}\left\{ 
\frac{1}{\Lambda ^{2}}\left[ \phi _{3}(\Lambda ^{2})-\phi _{3}(0)\right]
\right\} \right\} \frac{i\sigma _{\mu \nu }q^{\nu }}{2M}  \nonumber
\end{eqnarray}

\begin{eqnarray}
gsg &=&-ie^{2}\kappa F(q^{2})\left\{ -q^{2}S^{m-1}\left\{ \frac{1}{\Lambda
^{2}}\left[ \phi _{2}(\Lambda ^{2})-\phi _{2}(0)\right] \right\} \right\}
\gamma _{\mu }  \label{gsg2} \\
&&-ie^{2}\kappa F(q^{2})\left\{ 2(2M^{2}-q^{2})\left[ 
\begin{array}{c}
-N_{1}\phi _{1}(\lambda ^{2})+S^{m-1}\left\{ \frac{1}{\Lambda ^{2}}\phi
_{1}(\Lambda ^{2})\right\} \\ 
-S^{m-1}\left\{ \frac{1}{\Lambda ^{2}}\left[ \phi _{2}(\Lambda ^{2})-\phi
_{2}(0)\right] \right\}
\end{array}
\right] \right\} \frac{i\sigma _{\mu \nu }q^{\nu }}{2M}  \nonumber
\end{eqnarray}
\begin{eqnarray}
ggs+sgg &=&-2ie^{2}\kappa F(q^{2})\left\{ 
\begin{array}{c}
(\frac{q^{2}}{4}-3M^{2})S^{m-1}\left\{ \frac{1}{\Lambda ^{2}}\left[ \phi
_{3}(\Lambda ^{2})-\phi _{3}(0)\right] \right\} \\ 
+\frac{3}{2}S^{m-1}\left\{ \phi _{1}(\Lambda ^{2})-\frac{1}{2}\phi
_{2}(\Lambda ^{2})\right\}
\end{array}
\right\} \gamma _{\mu } \\
&&+-2ie^{2}\kappa F(q^{2})\left\{ 
\begin{array}{c}
2M^{2}S^{m-1}\left\{ \frac{1}{\Lambda ^{2}}\left[ \phi _{3}(\Lambda
^{2})-\phi _{3}(0)\right] \right\} \\ 
-4S^{m-1}\left\{ \phi _{1}(\Lambda ^{2})-\frac{1}{2}\phi _{2}(\Lambda
^{2})\right\}
\end{array}
\right\} \frac{i\sigma _{\mu \nu }q^{\nu }}{2M}  \nonumber
\end{eqnarray}
\begin{eqnarray}
gss+ssg &=&-2ie^{2}\kappa ^{2}F(q^{2})\frac{q^{2}}{4M^{2}}\left\{ 
\begin{array}{c}
-2M^{2}S^{m-1}\left\{ \frac{1}{\Lambda ^{2}}\left[ \phi _{3}(\Lambda
^{2})-\phi _{3}(0)\right] \right\} \\ 
-2S^{m-1}\left\{ \phi _{1}(\Lambda ^{2})-\frac{1}{2}\phi _{2}(\Lambda
^{2})\right\}
\end{array}
\right\} \gamma _{\mu } \\
&&+-2ie^{2}\kappa ^{2}F(q^{2})\left\{ 
\begin{array}{c}
(\frac{3}{4}q^{2}-M^{2})S^{m-1}\left\{ \frac{1}{\Lambda ^{2}}\left[ \phi
_{3}(\Lambda ^{2})-\phi _{3}(0)\right] \right\} \\ 
-\frac{1}{2}S^{m-1}\left\{ \phi _{1}(\Lambda ^{2})-\frac{1}{2}\phi
_{2}(\Lambda ^{2})\right\}
\end{array}
\right\} \frac{i\sigma _{\mu \nu }q^{\nu }}{2M}  \nonumber
\end{eqnarray}
\begin{eqnarray}
sgs &=&-ie^{2}\kappa ^{2}F(q^{2})\left\{ 
\begin{array}{c}
-2M^{2}S^{m-1}\left\{ \frac{1}{\Lambda ^{2}}\left[ \phi _{3}(\Lambda
^{2})-\phi _{3}(0)\right] \right\} -\frac{1}{2}S^{m-1}\left\{ \phi
_{3}(\Lambda ^{2})\right\} \\ 
+S^{m-1}\left\{ \phi _{1}(\Lambda ^{2})\right\} +\frac{1}{2}S^{m-1}\left\{
\phi _{2}(\Lambda ^{2})\right\} \\ 
-\frac{1}{2M^{2}}S^{m-1}\left\{ \Lambda ^{2}\left[ \phi _{1}(\Lambda ^{2})-%
\frac{1}{4}\phi _{2}(\Lambda ^{2})\right] \right\}
\end{array}
\right\} \gamma _{\mu } \\
&&-ie^{2}\kappa ^{2}F(q^{2})\left\{ 
\begin{array}{c}
2M^{2}S^{m-1}\left\{ \frac{1}{\Lambda ^{2}}\left[ \phi _{3}(\Lambda
^{2})-\phi _{3}(0)\right] \right\} \\ 
-2S^{m-1}\left\{ \phi _{1}(\Lambda ^{2})\right\} +\frac{1}{2}S^{m-1}\left\{
\phi _{3}(\Lambda ^{2})\right\}
\end{array}
\right\} \frac{i\sigma _{\mu \nu }q^{\nu }}{2M}  \nonumber
\end{eqnarray}

\begin{eqnarray}
sss &=&-ie^{2}\kappa ^{3}F(q^{2})\frac{q^{2}}{4M^{2}}\left\{ 
\begin{array}{c}
-2M^{2}S^{m-1}\left\{ \frac{1}{\Lambda ^{2}}\left[ \phi _{3}(\Lambda
^{2})-\phi _{3}(0)\right] \right\} +\frac{1}{2}S^{m-1}\left\{ \phi
_{3}(\Lambda ^{2})\right\} \\ 
+2S^{m-1}\left\{ \phi _{1}(\Lambda ^{2})-\phi _{2}(\Lambda ^{2})\right\}
\end{array}
\right\} \gamma _{\mu } \\
&&-ie^{2}\kappa ^{3}F(q^{2})\left\{ 
\begin{array}{c}
\frac{1}{2}q^{2}S^{m-1}\left\{ \frac{1}{\Lambda ^{2}}\left[ \phi
_{3}(\Lambda ^{2})-\phi _{3}(0)\right] \right\} -(1+\frac{q^{2}}{2M^{2}}%
)S^{m-1}\left\{ \phi _{1}(\Lambda ^{2})\right\} \\ 
-\frac{1}{2}S^{m-1}\left\{ \phi _{3}(\Lambda ^{2})\right\} \\ 
+\frac{1}{2}(1+\frac{q^{2}}{M^{2}})S^{m-1}\left\{ \phi _{2}(\Lambda
^{2})\right\} -\frac{1}{4M^{2}}S^{m-1}\left\{ \Lambda ^{2}\left[ \phi
_{1}(\Lambda ^{2})-\phi _{2}(\Lambda ^{2})\right] \right\}
\end{array}
\right\} \frac{i\sigma _{\mu \nu }q^{\nu }}{2M}  \nonumber
\end{eqnarray}
where 
\begin{equation}
S^{m-1}=N_{m}(\Lambda ^{2})^{m}T^{m-1}  \label{Sm1}
\end{equation}

It should be noted that the terms with $\phi _{1}(\lambda ^{2})$, which
appear only in $ggg$ and $gsg$, constitute the well-known infrared
divergence. They are, apart from the hard-photon proton interaction, (\ref
{cur}), independent of the proton form factor (in this case independent of $%
\Lambda $ and $M$). This is to be expected, since this term is cancelled by
a similar infrared divergent term coming from the cross section for the
emission of a real soft photon, which is given by the elastic cross section
multiplied by a factor independent of the proton form factor.

\smallskip

\section{Electron vertex correction}

As noted in Sec. IIIC, the electron vertex correction, $M_{5}$, may be
obtained directly from the proton vertex correction, $M_{6}$, if we retain
only the term $ggg$, set $F=1$, replace $p_{2},p_{4}$ and $M$ by $%
p_{1},p_{3} $ and $m$ and take the limit $\Lambda \rightarrow \infty $.
After making these replacements in $ggg$ as given in (\ref{ggg2}), we need $%
\phi _{k}(\Lambda ^{2})$ to order $m^{2}/\Lambda ^{2}$ and $\phi _{k}(0)$
(see (\ref{B36})). From (\ref{phi1}), writing (\cite{thooft}, p. 389 (B.2)) 
\[
L(1-z)=-L(z)+\frac{1}{6}\pi ^{2}-\ln z\,\ln (1-z) 
\]
and neglecting terms of relative order $m^{2}/\Lambda ^{2}$, we have 
\begin{equation}
\phi _{1}(\Lambda ^{2})\,\,_{\overrightarrow{\Lambda \rightarrow \infty }%
}\,\,\frac{1}{\Lambda ^{2}}\left\{ 1+\ln \left( \frac{\Lambda ^{2}}{m^{2}}%
\right) -\frac{\rho _{m}}{\rho _{1}}\ln x_{m}\right\}
\end{equation}
and from (\ref{phik}) 
\begin{equation}
\phi _{k}(\Lambda ^{2})-\phi _{k+1}(\Lambda ^{2})\,\,_{\overrightarrow{%
\Lambda \rightarrow \infty }}\,\frac{1}{\Lambda ^{2}}\left\{ \frac{1}{k+1}%
\right\} \,
\end{equation}
Choosing $m=1$ in (\ref{ggg2}) we then have 
\begin{equation}
ggg=\left[ G_{1}^{(e)}(q^{2})\gamma _{\mu }+G_{2}^{(e)}(q^{2})\frac{i\sigma
_{\mu \nu }q^{\nu }}{2m}\right]
\end{equation}
where 
\begin{equation}
G_{1}^{(e)}(q^{2})=\frac{\alpha }{4\pi }\left\{ -2(2m^{2}-q^{2})\phi
_{1}(\lambda ^{2})+\left( \frac{3\rho _{m}^{2}-4m^{2}}{\rho _{m}\rho _{1}}%
\right) \ln x_{m}+\frac{1}{2}+\ln \left( \frac{\Lambda ^{2}}{m^{2}}\right)
\right\}
\end{equation}
\begin{equation}
G_{2}^{(e)}(q^{2})=\frac{\alpha }{4\pi }\left\{ \frac{4m^{2}}{\rho _{m}\rho
_{1}}\ln x_{m}\right\}
\end{equation}
Adding the contribution of the electron self energy diagrams gives 
\begin{equation}
\overline{ggg}=\left[ \left( G_{1}^{(e)}(q^{2})-G_{1}^{(e)}(0)\right) \gamma
_{\mu }+G_{2}^{(e)}(q^{2})\frac{i\sigma _{\mu \nu }q^{\nu }}{2m}\right]
\end{equation}
To facilitate comparison with \cite{tsai}, we write this in terms of the
functions $K(p_{i},p_{j})$. Similarly to (\ref{Kp2p4}), we now have 
\begin{equation}
(2m^{2}-q^{2})\phi _{1}(\lambda ^{2})=K(p_{1},p_{3})
\end{equation}
which then gives (\ref{elvtx}).

\section{The functions \protect\boldmath{$\phi _{\lowercase{k}}\left(
\Lambda ^{2}\right) $}}

In this Appendix we derive the three-term recurrence relation for the
function $\phi _{k}(\Lambda ^{2})$ given in (\ref{recur}) as well as the
expression for the function $\phi _{1}(\Lambda ^{2})$, defined in (\ref{phik}%
) and given in terms of Spence functions in (\ref{phi1}). Integrating first
over $y$ in Eq. (\ref{phik}) we have

\begin{equation}
\phi _{k}(\Lambda ^{2})=\frac{1}{\rho _{1}}\int_{0}^{1}\frac{x^{k-1}}{R}\log
\left\{ \frac{R+x\rho _{1}}{R-x\rho _{1}}\right\} \,dx,
\end{equation}
where $R^{2}=\rho ^{2}x^{2}+4(1-x)\Lambda ^{2}$ and $\rho _{1}^{2}=-q^{2}>0$%
. Noting that

\begin{equation}
\frac{d}{dx}\left\{ x^{k}R\right\} =x^{k-1}\left\{ kR^{2}+\rho
^{2}x^{2}-2x\Lambda ^{2}\right\} R^{-1},
\end{equation}
we get 
\begin{equation}
(k+1)\rho ^{2}\phi _{k+2}-2(2k+1)\Lambda ^{2}\phi _{k+1}+4k\Lambda ^{2}\phi
_{k}=\frac{2}{\rho _{1}}\int_{0}^{1}\log \left\{ \frac{R+x\rho _{1}}{R-x\rho
_{1}}\right\} \,d\left( x^{k}R\right) .
\end{equation}
Integration by parts then gives 
\begin{equation}
(k+1)\rho ^{2}\phi _{k+2}-2(2k+1)\Lambda ^{2}\phi _{k+1}+4k\Lambda ^{2}\phi
_{k}=\frac{2\rho }{\rho _{1}}\log \left( \frac{\rho +\rho _{1}}{\rho -\rho
_{1}}\right) -2\Lambda ^{2}\int_{0}^{1}\frac{x^{k}(2-x)}{M^{2}x^{2}+(1-x)%
\Lambda ^{2}}  \label{D5}
\end{equation}
from which the recursion (\ref{recur}) follows at once, using Eq. (\ref{B28p}%
)

From Eq. (\ref{D5}) it is clear that $\phi _{2}$ and $\phi _{3}$ follow once
we have evaluated $\phi _{1}$, which follows. Setting $k=1$ in (\ref{phik})
and integrating first over $x$, we get 
\begin{equation}
\int_{0}^{1}\,\frac{x\,dx}{p_{y}^{2}x^{2}+\Lambda ^{2}(1-x)}=\frac{1}{%
2p_{y}^{2}}\left[ \ln \left( \frac{p_{y}^{2}}{\Lambda ^{2}}\right) +\frac{%
\Lambda }{\Delta }\ln \left( \frac{\Lambda +\Delta }{\Lambda -\Delta }%
\right) \right] ,
\end{equation}
where $\Delta ^{2}=\Lambda ^{2}-4p_{y}^{2}$. We next make the change of
variable $y=(1+\omega )/2$, which gives $\Delta ^{2}=\rho _{1}^{2}\omega
^{2}+\Lambda ^{2}-\rho ^{2}$, and then make the further change of variable $%
\Delta =\rho _{1}\omega +s$, from which 
\begin{equation}
\omega =\frac{\Lambda ^{2}-\rho ^{2}-s^{2}}{2\rho _{1}s};\,\,\,\,\,\,\,\,\,%
\,\,\,\Delta =\frac{\Lambda ^{2}-\rho ^{2}+s^{2}}{2s}
\end{equation}
Then integrating (\ref{phik}) over $y$ gives 
\begin{eqnarray}
\phi _{1}(\Lambda ^{2}) &=&\frac{2}{\rho _{1}}\int_{s_{-}}^{s_{+}}\frac{ds}{%
\rho ^{2}-(\Lambda -s)^{2}}\ln \left[ \frac{(\Lambda +s)^{2}-\rho ^{2}}{%
4s\Lambda }\right] \\
&&-\frac{2}{\rho _{1}}\int_{s_{-}}^{s_{+}}\frac{ds}{(\Lambda +s)^{2}-\rho
^{2}}\ln \left[ \frac{\rho ^{2}-(\Lambda -s)^{2}}{4s\Lambda }\right] 
\nonumber
\end{eqnarray}
where 
\begin{equation}
s_{\pm }=\Lambda _{1}\pm \rho _{1}
\end{equation}
Factoring the expressions which appear as factors to the logarithms as well
as in their arguments, we can write 
\begin{equation}
\phi _{1}(\Lambda ^{2})=\frac{1}{\rho \rho _{1}}\sum_{i=1}^{9}I_{i}
\label{D10}
\end{equation}
where 
\[
I_{1}=\int_{s_{-}}^{s_{+}}\frac{ds}{s-\sigma _{-}}\ln (s+\sigma _{+});\text{ 
}I_{2}=\int_{s_{-}}^{s_{+}}\frac{ds}{s+\sigma _{+}}\ln (s-\sigma _{-}) 
\]
\[
I_{3}=\int_{s_{-}}^{s_{+}}\frac{ds}{\sigma _{+}-s}\ln (s+\sigma _{-});\text{ 
}I_{4}=-\int_{s_{-}}^{s_{+}}\frac{ds}{s+\sigma _{-}}\ln (\sigma _{+}-s) 
\]
\begin{equation}
I_{5}=\int_{s_{-}}^{s_{+}}\frac{ds}{s-\sigma _{-}}\ln \left( \frac{s+\sigma
_{-}}{2s}\right) ;\text{ }I_{8}=-\int_{s_{-}}^{s_{+}}\frac{ds}{s+\sigma _{-}}%
\ln \left( \frac{s-\sigma _{-}}{2s}\right)
\end{equation}
\[
I_{6}=\int_{s_{-}}^{s_{+}}\frac{ds}{s+\sigma _{+}}\ln \left( \frac{\sigma
_{+}-s}{2s}\right) ;\text{ }I_{7}=\int_{s_{-}}^{s_{+}}\frac{ds}{\sigma _{+}-s%
}\ln \left( \frac{s+\sigma _{+}}{2s}\right) 
\]
\[
I_{9}=-\int_{s_{-}}^{s_{+}}ds\left[ \frac{1}{s-\sigma _{-}}+\frac{1}{\sigma
_{+}-s}-\frac{1}{\bar{s}+\sigma _{-}}+\frac{1}{s+\sigma _{+}}\right] \ln
(2\Lambda ) 
\]
where 
\begin{equation}
\sigma _{\pm }=\Lambda \pm \rho
\end{equation}
Some of the integrals $I_{i}$ can be integrated out directly. For these
terms we get: 
\begin{equation}
\sum_{n=1}^{4}I_{n}=\ln \left( \frac{\alpha _{-}}{\alpha _{+}}\right)
ln\left( \frac{4\sigma _{+}\sigma _{-}}{(1-\alpha _{+}^{2})(1-\alpha
_{-}^{2})}\right) -\ln \alpha _{+}\ln (1-\alpha _{+}^{2})+\ln \alpha _{-}\ln
(1-\alpha _{-}^{2})
\end{equation}
and 
\begin{equation}
I_{9}=-2\ln \left( \frac{\alpha _{-}}{\alpha _{+}}\right) \ln (2\Lambda )
\end{equation}
where 
\[
\alpha _{\pm }=\frac{\rho \mp \rho _{1}}{\Lambda +\Lambda _{1}} 
\]
The remaining terms in Eq. (\ref{D10}) we may rewrite as 
\begin{equation}
\sum_{n=5}^{8}I_{n}=\ln \alpha _{+}\ln (1-\alpha _{+}^{2})-\ln \alpha
_{-}\ln (1-\alpha _{-}^{2})+L(1-\alpha _{+}^{2})-L(1-\alpha _{-}^{2})
\end{equation}
where $L$ is the dilogarithm (Spence) function. We then get the result given
in (\ref{phi1}) in Appendix B.

\section{Final electron detector acceptance}

In this Appendix, we express $\Delta \epsilon $, the maximum momentum of the
photon in the frame $S^{0}$, in terms of the final electron detector
acceptance in the lab frame, $\Delta E$. \thinspace In $S^{0}$ ({\bf p}$%
_{4}+ $ {\bf k} $=0$), if $|${\bf k$|$} $=\Delta \epsilon \ll M$, we have,
from $(p_{1}+p_{2}-p_{3})^{2}=(p_{4}+k)^{2},$ neglecting terms of order $%
(\Delta \epsilon /M)^{2}$ and $(m/M)^{2}$, 
\begin{equation}
p_{2}\cdot (p_{1}-p_{3})-p_{1}\cdot p_{3}=M\Delta \epsilon
\end{equation}
Writing this in terms of lab frame energies, we have, for high energies, 
\begin{equation}
M(\epsilon _{1}-\epsilon _{3})-\epsilon _{1}\epsilon _{3}(1-\cos \theta
)=M\Delta \epsilon
\end{equation}
For elastic scattering in the lab frame, we have 
\begin{equation}
M(\epsilon _{1}-\epsilon _{3}^{el})-\epsilon _{1}\epsilon _{3}^{el}(1-\cos
\theta )=0
\end{equation}
Subtracting gives 
\begin{equation}
\Delta E\left( 1+\frac{\epsilon _{1}}{M}(1-\cos \theta )\right) =\Delta
\epsilon
\end{equation}
where 
\begin{equation}
\Delta E=\epsilon _{3}^{el}-\epsilon _{3}
\end{equation}
Thus, in terms of lab frame quantities we have 
\begin{equation}
\Delta \epsilon =\eta \Delta E
\end{equation}

\section{High energy approximation for $S_{\lowercase{ij}}^{(2)}$}

In this appendix we give the high energy approximation of the terms $%
S_{ij}^{(2)}$ defined in (\ref{4.15}), in which we note in particular that
for $i=1$ or $3$ we have $l\cdot t=(\alpha p_{i}-p_{j})\cdot t\approx \alpha
p_{i}\cdot t$. Using transformations of the dilog (Spence) functions \cite
{thooft}, p. 389 (B.3), 
\begin{equation}
L(z)=-L\left( \frac{1}{z}\right) -\frac{1}{6}\pi ^{2}-\frac{1}{2}\ln ^{2}(-z)
\end{equation}
\begin{equation}
L(z)=-L\left( \frac{z}{z-1}\right) -\frac{1}{2}\ln ^{2}\left( 1-z\right)
\end{equation}
the terms in $S_{ij}^{(2)}$ simplify considerably. We then obtain 
\begin{eqnarray}
S_{12}^{(2)} &=&-\ln ^{2}\left( \frac{2\epsilon _{3}}{m}\right) -\ln ^{2}x+%
\frac{1}{2}\ln ^{2}\left( \frac{x}{\eta }\right) -\frac{1}{6}\pi ^{2} \\
&&-L\left( 1-\frac{1}{x\eta }\right) +L\left( 1-\frac{\eta }{x}\right) 
\nonumber
\end{eqnarray}

\begin{eqnarray}
S_{32}^{(2)} &=&-\ln ^{2}\left( \frac{2\epsilon _{1}}{m}\right) -\ln
^{2}\left( x\right) +\frac{1}{2}\ln ^{2}\left( x\eta \right) -\frac{1}{6}\pi
^{2} \\
&&+L\left( 1-\frac{1}{x\eta }\right) -L\left( 1-\frac{\eta }{x}\right) 
\nonumber
\end{eqnarray}
\begin{equation}
S_{14}^{(2)}=-\ln ^{2}\left( \frac{2\epsilon _{3}}{m}\right) -\frac{1}{6}\pi
^{2}
\end{equation}
\begin{equation}
S_{34}^{(2)}=-\ln ^{2}\left( \frac{2\epsilon _{1}}{m}\right) -\frac{1}{6}\pi
^{2}
\end{equation}
\begin{eqnarray}
S_{13}^{(2)} &=&-\ln ^{2}\left( \frac{2\epsilon _{1}}{m}\right) -\ln
^{2}\left( \frac{2\epsilon _{3}}{m}\right) -\frac{1}{3}\pi ^{2}+\frac{1}{2}%
\ln ^{2}\left( \cos ^{2}\frac{1}{2}\theta \right) \\
&&+L\left( \cos ^{2}\frac{1}{2}\theta \right)  \nonumber
\end{eqnarray}
\begin{equation}
S_{24}^{(2)}=\frac{1}{2}\ln ^{2}\left( x\right) +\frac{1}{2}L\left( 1-\frac{1%
}{x^{2}}\right)
\end{equation}

\begin{figure}[tbp]
\caption{Feynman diagrams for elastic amplitudes.}
\label{fig1}
\end{figure}

\begin{figure}[tbp]
\caption{Feynman diagrams for inelastic amplitudes.}
\label{fig2}
\end{figure}

\begin{figure}[tbp]
\caption{The curves show the contribution of nucleonic size effects (VTX -
dashed curve), mathematical refinement (D0 - dotted curve), and the
resulting difference (D - solid curve) between the radiative correction
given by Mo and Tsai \protect\cite{mo} and that given in this paper, $%
\delta_{\text{MTj}}$, as a function of four-momentum transfer for an initial
electtron energy of 6 GeV. (VTX, D0, and D are defined in Sec. V.)}
\label{fig3}
\end{figure}

\begin{figure}[tbp]
\caption{As Fig. \ref{fig3}, but with an initial electron energy of 16 GeV.}
\label{fig4}
\end{figure}

\begin{table}[tbp]
\caption{Contributions to the radiative correction $\delta$ for
electron-proton scattering as given in this paper (MTj) and in Mo and Tsai 
\protect\cite{mo} (MoTsai) for three initial electron energies and
four-momentum transfers. Values in the rows marked $Z^{0}, Z^{1}$, and $%
Z^{2} $ refer to contributions from terms with these factors in (\ref{delta}%
).}
\label{table1}
\begin{tabular}{lllllll}
& \multicolumn{2}{l}{$\epsilon _{1}=4.4$ GeV} & \multicolumn{2}{l}{$\epsilon
_{1}=12$ GeV} & \multicolumn{2}{l}{$\epsilon _{1}=21.5$ GeV $\,$} \\ 
& \multicolumn{2}{l}{$Q^{2}=$ 6 (GeV/$c$)$^{2}$} & \multicolumn{2}{l}{$%
Q^{2}= $ 16 (GeV/$c$)$^{2}$} & \multicolumn{2}{l}{$Q^{2}=$ 31.3 (GeV/$c$)$%
^{2} $} \\ \hline
& MTj & MoTsai & MTj & MoTsai & MTj & MoTsai \\ \hline
$Z^{0}$ & $-$0.2187 & $-$0.2171 & $-$0.2330 & $-$0.2322 & $-$0.2323 & $-$%
0.2317 \\ 
$Z^{1}$ & $-$0.0569 & $-$0.0506 & $-$0.0517 & $-$0.0479 & $-$0.0625 & $-$%
0.0571 \\ 
$Z^{2}$ & $-$0.0242 & $-$0.0232 & $-$0.0359 & $-$0.0347 & $-$0.0452 & $-$%
0.0440 \\ 
$\delta _{el}^{(1)}$ & $+$0.0068 &  & $+$0.0116 &  & $+$0.0185 &  \\ 
$\delta $ & $-$0.2930 & $-$0.2908 & $-$0.3090 & $-$0.3149 & $-$0.3214 & $-$%
0.3328
\end{tabular}
\end{table}

\begin{table}[tbp]
\caption{Contributions to the radiative correction $\delta$ as given in this
paper (MTj) and in Mo and Tsai \protect\cite{mo} (MoTsai) for several
nuclei, with $\epsilon _{1}=4.4$ GeV, $\,\,Q^{2}=$ 6 (GeV/$c$)$^{2}$; other
symbols as in Table \ref{table1}.}
\label{table2}%
\begin{tabular}{ddddddddd}
& \multicolumn{2}{c}{$^{2}$H} & \multicolumn{2}{c}{$^{4}$He} &
\multicolumn{2}{c}{$^{12}$C} & \multicolumn{2}{c}{$^{40}$Ca} \\ \hline
& MTj & MoTsai & MTj & MoTsai & MTj & MoTsai & MTj & MoTsai \\ \hline
$Z^{0}$ & $-$0.2476 & $-$0.2467 & $-$0.2535 & $-$0.2532 & $-$0.2615 & $-$0.2609 &
$-$0.2632 & $-$0.2625 \\
$Z^{1}$ & $-$0.0187 & $-$0.0183 & $-$0.0066 & $-$0.0077 & $-$0.0151 & $-$0.0168 &
$-$0.0147 & $-$0.0173 \\
$Z^{2}$ & $-$0.0094 & $-$0.0088 & $-$0.0077 & $-$0.0071 & $-$0.0156 & $-$0.0145 &
$-$0.0188 & $-$0.0178 \\
$\delta _{el}^{(1)}$ & $+$0.0010 &  & $+$0.0002 &  & $+$0.0001 &  & $+$0.0001 & \\
$\delta $ & $-$0.2747 & $-$0.2739 & $-$0.2677 & $-$0.2680 & $-$0.2920 & $-$0.2922
&$-$0.2966 & $-$0.2975
\end{tabular}
\end{table}

\begin{table}[tbp]
\caption{Contributions to the radiative correction $\delta$ as given in this
paper (MTj) and in Mo and Tsai \protect\cite{mo} (MoTsai) for several
nuclei, with $\epsilon _{1}=21.5$ GeV, $\,Q^{2}=$ 31.3 (GeV/$c$)$^{2}$ ;
other symbols as in Table \ref{table1}.}
\label{table3}
\begin{tabular}{lllllllll}
& \multicolumn{2}{c}{$^{2}$H} & \multicolumn{2}{c}{$^{4}$He} & 
\multicolumn{2}{c}{$^{12}$C} & \multicolumn{2}{c}{$^{40}$Ca} \\ \hline
& MTj & MoTsai & MTj & MoTsai & MTj & MoTsai & MTj & MoTsai \\ \hline
$Z^{0}$ & $-$0.2707 & $-$0.2704 & $-$0.2817 & $-$0.2814 & $-$0.2876 & $-$%
0.2875 & $-$0.2896 & $-$0.2894 \\ 
$Z^{1}$ & $-$0.0182 & $-$0.0189 & $-$0.0149 & $-$0.0166 & $-$0.0134 & $-$%
0.0164 & $-$0.0123 & $-$0.0173 \\ 
$Z^{2}$ & $-$0.0231 & $-$0.0221 & $-$0.0379 & $-$0.0352 & $-$0.0583 & $-$%
0.0535 & $-$0.0761 & $-$0.0708 \\ 
$\delta _{el}^{(1)}$ & $+$0.0045 &  & $+$0.0026 &  & $+$0.0006 &  & $+$0.0001
&  \\ 
$\delta $ & $-$0.3076 & $-$0.3114 & $-$0.3319 & $-$0.3333 & $-$0.3587 & $-$%
0.3573 & $-$0.3786 & $-$0.3775
\end{tabular}
\end{table}

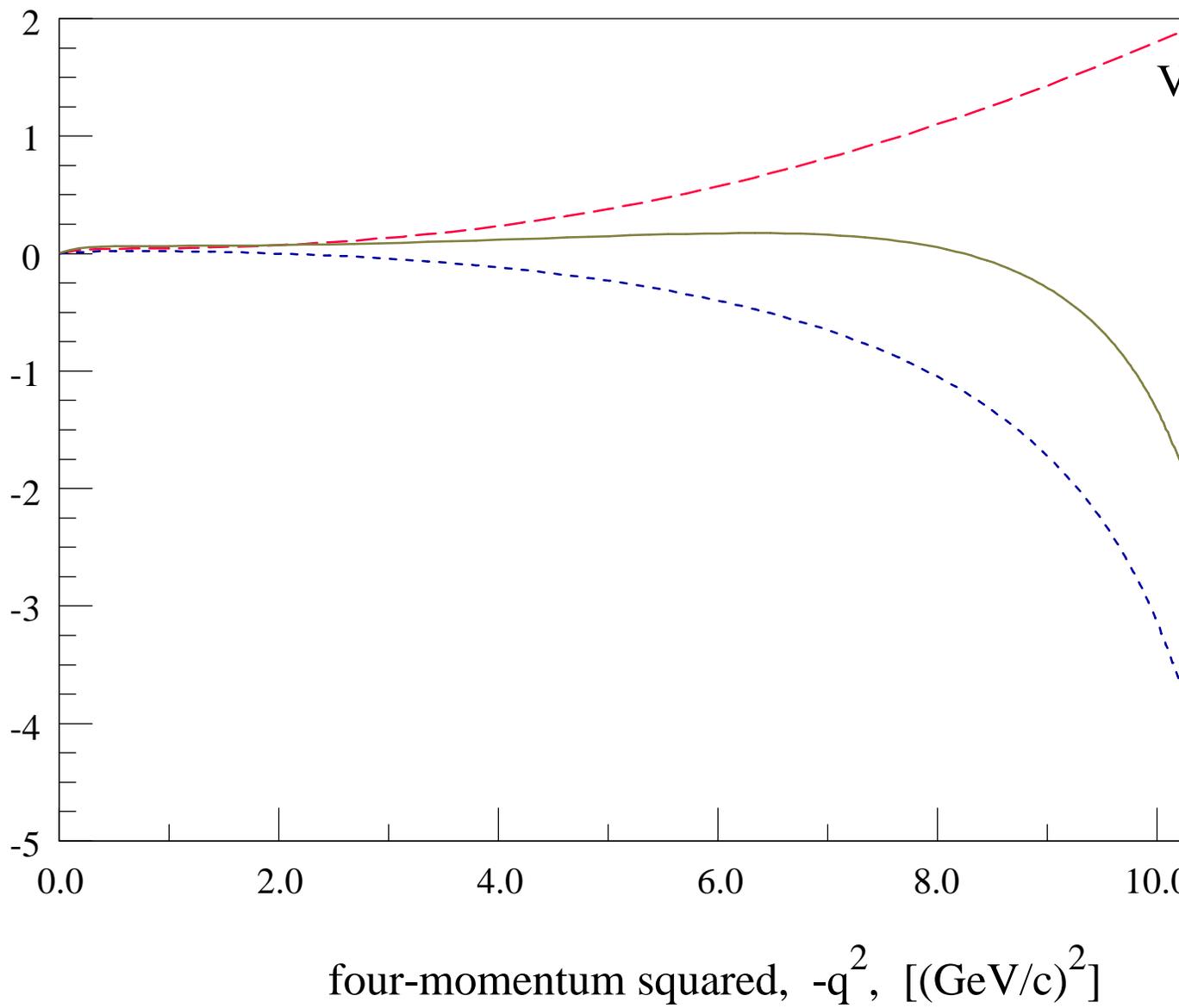
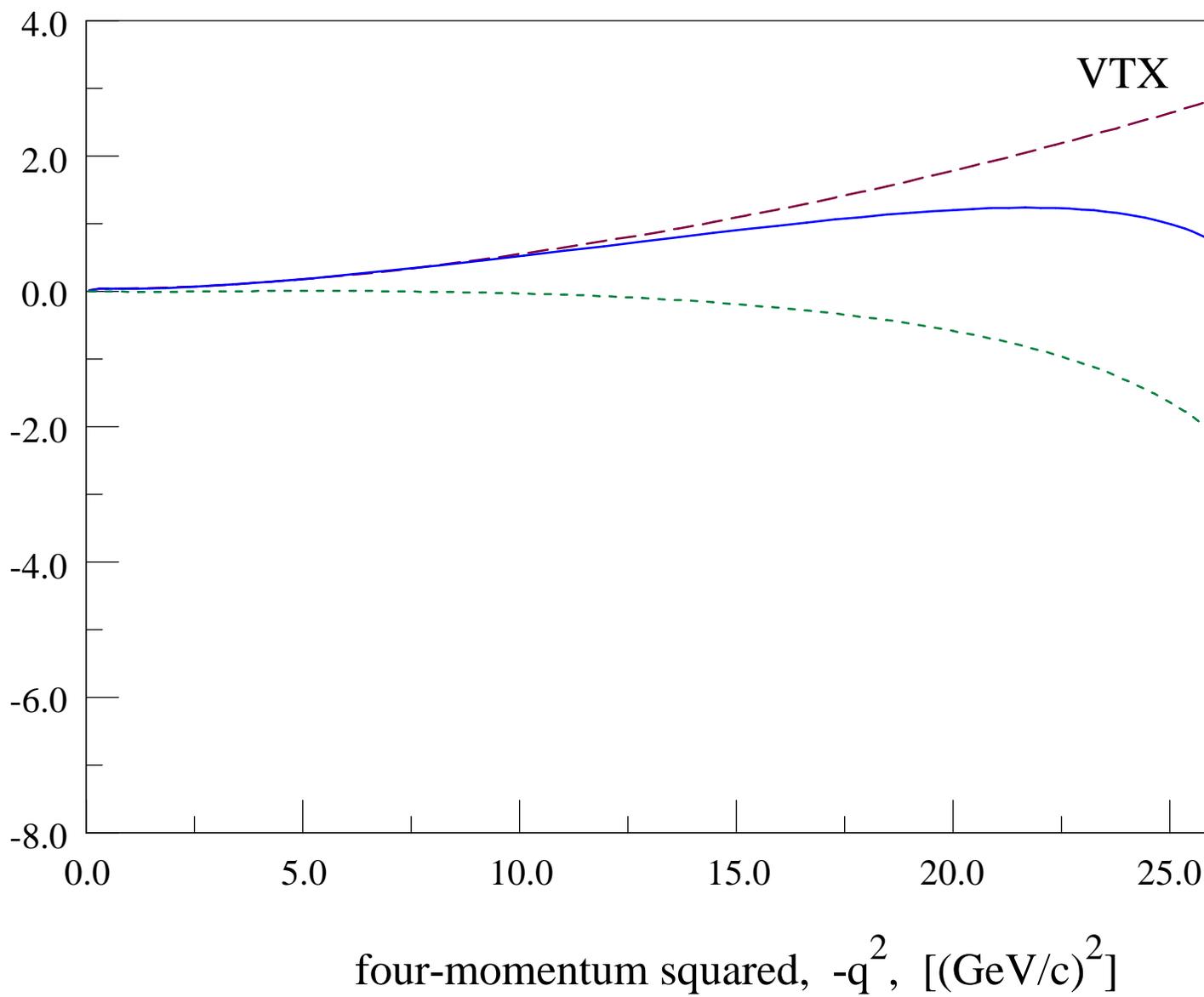
\end{document}